\documentstyle[12pt,epsf]{article}
\def\e{\varepsilon}

\def\be{\begin{equation}}
\def\ee{\end{equation}}
\def\ba{\begin{eqnarray}}
\def\ea{\end{eqnarray}}
\def\a{\alpha}
\def\b{\beta}

\def\th{\vartheta}

\def\l{\lambda}

\def\im{{\rm Im}\tau}

\font\fivesans=cmss10 at 4.61pt
\font\sevensans=cmss10 at 6.81pt
\font\tensans=cmss10
\newfam\sansfam
\textfont\sansfam=\tensans\scriptfont\sansfam=
\sevensans\scriptscriptfont
\sansfam=\fivesans
\def\sans{\fam\sansfam\tensans}

\def\Z{{\mathchoice
{\hbox{$\sans\textstyle Z\kern-0.4em Z$}}
{\hbox{$\sans\textstyle Z\kern-0.4em Z$}}
{\hbox{$\sans\scriptstyle Z\kern-0.3em Z$}}
{\hbox{$\sans\scriptscriptstyle Z\kern-0.2em Z$}}}}
\newlength{\boxsize} %From Ken

\def\lrp{\stackrel{\leftrightarrow}{\partial}}

\begin{document}
\begin{flushright}
{CERN-TH/97-57\\
 LPTENS/97/16\\
 hep-th/9703198}
\end{flushright}
\vspace*{0.5cm}
\begin{center}

{\bf BPS STATES IN SUPERSTRINGS WITH SPONTANEOUSLY BROKEN SUSY}\\
\vspace*{1.0cm}
Costas Kounnas$^{*)}$

Theory Division, CERN\\
CH-1211 Geneva 23, Switzerland

\vspace*{1.0cm}
Abstract
\vspace*{1.0cm}
\end{center}
We show the existence of a supersymmetry-breaking mechanism in
string theory, where $N=4$ supersymmetry is broken
{\em spontaneously} to $N=2$  and $N=1$ with moduli-dependent
gravitino masses. The BPS spectrum of the theory with lower
supersymmetry is in one-to-one correspondence with the spectrum of
the heterotic $N=4$ string. The mass splitting of the $N=4$ spectrum
depends on the  moduli as well as the three $R$-symmetry charges.
In the case of $N=4 \rightarrow N=2$, the perturbative $N=2$
prepotential is determined by the perturbative $N=4$ BPS states. This
observation led us to  suggest a  method that determines the exact
non-perturbative prepotential of the effective $N=2$ supergravity
using the shifted spectrum  of the  non-perturbative BPS states of
the underlying $N=4$ theory.
\vspace*{1.5cm}
\begin{center}
{\it Summary of Lectures given at Summer School}\\
{\it Ecole de Carg\`ese, France, 5--17 August 1996}\\
{\it and at the}\\
{\it Conference on ``Advanced Quantum Field Theory",}\\
{\it in memory of Claude Itzykson}\\
{\it La Londe-les-Maures, France, 31 August -- 5 September 1996}\\
\end{center}
\vspace*{2.0cm}
\rule[.1in]{16.5cm}{.002in}

\noindent
$^{*)}$  On leave from Ecole Normale Sup\'erieure, 24 rue Lhomond, F-75231
Paris, Cedex 05, France, {\it e-mail:  kounnas@nxth04.cern.ch}.
\vspace*{1.0cm}
\begin{flushleft}
CERN-TH/97-57\\
March 1997
\end{flushleft}
\vfill\eject

\setcounter{page}{1}
%\pagestyle{plain}
%\numberwithin{equation}{section}
%\renewcommand{\theequation}{\thesection\arabic{equation}}
\renewcommand{\theequation}{\arabic{section}.\arabic{equation}}

\section{Introduction}
\setcounter{equation}{0}

When a local symmetry is spontaneously broken, the physical states
can
be classified in terms of the unbroken phase spectrum and in terms of
a well-defined mass splitting given in terms of vacuum expectation
values of some fields,
weighted by the charges of the broken symmetry. In the case of
gauge symmetry breaking, the fields with non-zero vev's are physical
scalar fields, while in the case of supersymmetry breaking they are
auxiliary fields. In extended supersymmetric theories (local or
global), the supersymmetric vacua are  degenerate, with zero vacuum
energy for any vev of the moduli fields ($S,T^i$). For instance, in
the case of $N=4$ supergravity based on a gauge group
$U(1)^6\times~G$, the space of the moduli fields is given in terms of
    $2+6r$ physical scalars, which are coordinates  of the coset
space \cite{Roo},\cite{fgkp}

\begin{equation}
\left[{SU(1,1)\over U(1)}\right]_S\times \left[{SO(6, r)\over
{SO(6)\times SO(r)}}\right]_{T}.
\label{moduli}
\ee
$r$ is the rank of the gauge group $G$.

In an arbitrary point of the moduli space the gauge symmetry $G$ is
broken down to $U(1)^r$ while at some special points of the moduli
space the gauge symmetry
is extended to some non--Abelian gauge group of  the same rank due to
the presence of some extra gauge multiplets that become massless at
the  special points of the moduli space.

In the heterotic $N=4$ superstring solution obtained by $T^6$
compactification of the  $10-d$ superstring, the rank of the
group $r$ has a fixed value, $r=22$ 
\cite{nsw}-\cite{lls}. In an
arbitrary point of the
moduli space the gauge group is $U(1)^r$
and in special points the symmetry is extended as in field theory.
There is however a fundamental difference between the field theory
Higgs phenomenon and the string theory  one. Indeed, if in  an $N=4$
field theory the gauge group is $G=U(1)^6\times SO(32)$ at any given
point of the
moduli space, then at any other point the remaining gauge symmetry
$G_{T_i} $ is always a subgroup of  $G$ with smaller dimensionality
dim$(G_{T_i})\le$ dim($G$). On the contrary, in the string Higgs
phenomenon, owing to the existence of winding states, we can connect
gauge groups
which are not subgroups of a larger  gauge group. For instance, it is
possible to connect continuously $G= U(1)^6\times SO(32)$ with  $G=
U(1)^6\times E_8\times E_8$, as well as with the most symmetric of
same rank, namely  $G=SO(44)$.
Indeed, starting from a $10-d$ $N=1$ supergravity theory with
$G=SO(32)$ or $G=E_8\times E_8$ after compactification in four
dimensions the only possible $N=4$ supergravity effective theories
are
based either to $G=U(1)^6\times SO(32)$ or $G=U(1)^6\times E_8\times
E_8$ (and their subgroups obtained with Higgs phenomenon). In string
theory the gauge group can be further extended due to the existence
of extra gauge bosons with non-zero winding numbers, which can become
massless in special points of the moduli space.

When some auxiliary fields of the supergravity theories have
non--vanishing
vev, some (or all) of the supersymmetries are spontaneously broken
\cite{ss1}--\cite{kr}.
There is a consistent class of $N=1,2$ and $N=4$ models defined in
flat space-time in which all supersymmetries are broken or partially
    broken \cite{abk}--\cite{kr}. The most interesting case for our
purposes is that in which there is one of the
supersymmetries left unbroken. In that case we know that it is
possible, in general, to have chiral representations of matter scalar
multiplets
which can describe the quarks and leptons of the supersymmetric
standard model.  All previous examples about the
partial breaking of $N=2$ to $N=1$ supersymmetry was done at the
    field theory level \cite{apt}. In this work we will first show the
extension of the partial spontaneous breaking at the perturbative
string level and then we will generalize our result to the
non-perturbative level using as a tool the heterotic--type II string
duality of the $N=4$ 4d-superstrings \cite{HT},\cite{wi}.

\section{Perturbative  $N=4$ Mass Spectrum}
\setcounter{equation}{0}

Our starting point is a four-dimensional heterotic $N=4$ superstring
solution.
{}From  the world-sheet viewpoint, these theories are constructed by
the following left- and right-moving degrees of freedom:\\
$\bullet$ Four left-moving non-compact supercoordinates,
{\bf $X^{\mu}, \Psi^{\mu}$}\\
$\bullet$ Six left-moving compactified supercoordinates,
{\bf $\Phi^{I}, \Psi^{I}$}\\
$\bullet$ The left-moving super-ghosts,
{\bf $b, c$} and {\bf $\beta, \gamma$}\\
$\bullet$ Four right-moving  coordinates,
{\bf  ${\bar X}^{\mu}$}\\
$\bullet$ Six right-moving compactified coordinates,
{\bf  ${\bar \Phi}^{I}$}\\
$\bullet$ 32 right-moving fermions,
{\bf ${\bar \Psi}^A$}\\
$\bullet$ The right-moving ghosts,
{\bf $\bar b, \bar c$}
\vskip .2cm
In order to obtain a consistent (without ghosts) $N=4$ solution the
left-moving fermions
$\Psi^{\mu}, \Psi^{I}$ and the $\beta, \gamma$ ghosts must have the
same boundary conditions. In that case the global existence of the
left-moving spin-$3/2$ world-sheet supercurrent
\be
T_F~=~\Psi^{\mu}\partial X^{\mu} + \Psi^{I}\partial\Phi^I
\label{supcur}
\ee
implies periodic boundary conditions for the compact and non-compact
left-moving coordinates, $\Phi^I,~ X^{\mu}$. Modular invariance
implies the right-moving coordinates ${\bar \Phi}^I,~ {\bar
X}^{\mu}$ to be periodic as well.
The solution with $G=U(1)^6\times SO(32)$, is when the right-moving
fermions $\Psi^{A}$ have the same boundary conditions (periodic or
antiperiodic), while the solution with $G=U(1)^6\times E_8\times E_8$
is when the ${\bar\Psi}^{A}=({\bar \Psi}^{A_1},~ {\bar \Psi}^{A_2})$
are in two groups of sixteen with the same boundary conditions.
Starting either from the $G=U(1)^6\times E_8\times E_8$ solution  or
from the $G=U(1)^6\times SO(32)$  we can obtain all others by
deforming the momentum lattice of compactified bosons together with
the charge lattice of the 32 fermions ${\bar\Psi}^{A}$.

The  partition function of the heterotic $N=4$ solutions in a generic
point of the moduli space is well known and has the following
expression \cite{nsw}:
\ba
Z(\tau ,{\bar\tau})=
    {\im ^{-1}~\over {\eta^2(\tau)~
{\bar\eta}^2({\bar\tau})}}\times{1\over 2}\sum_{\alpha,\beta}
(-)^{\alpha+\beta +\alpha \beta}~
\frac{\th^{4}[^{\alpha}_{\beta}](\tau)}{\eta^4(\tau)}~
 {\Gamma_{(6,22)}(\tau,{\bar\tau})\over
\eta^6(\tau)\bar\eta^{22}(\bar \tau)}~~,
\label{n4}
\ea
where $ \Gamma_{(6,22)}(\tau,{\bar\tau})$ denotes the partition
function
due to the compactified coordinates $\Phi^I, {\bar \Phi}^A$ and due
to the sixteen right-moving $U(1)$ currents constructed with the
fermions ${\bar \Psi^I}$
\be
{\bar J}^k= {\bar \Psi}^{2k-1}{\bar \Psi}^{2k},~~~~k=1,2,...,16.
\label{b1}\ee
$\Gamma_{(6,22)}(\tau,{\bar\tau})$ has in total $6\times 22$ moduli
parameters which correspond to (1,1) marginal deformations
of the world-sheet action.
\be
\delta S^{2d}~=~\delta T_{IJ}~\partial \Phi^I~\partial {\bar \Phi}^J
{}~+~
Y_{I}^{k}~\partial \Phi^I~{\bar J}^k.
\label{b2}\ee

In terms of the six-dimensional backgrounds of the compactified
space, the $T_{IJ}$ moduli are related to the internal background
metric
$G_{IJ}$ and  the  internal antisymmetric tensor $B_{IJ}$; $T_{IJ}$
=$G_{IJ}$+ $B_{IJ}$. The $Y_{I}^{k}$ moduli are the six-dimensional
internal gauge field backgrounds which belong in the Cartan
subalgebra of the ten-dimensional gauge group (either $E_8\times
E_8$ or $SO(32)$). From the four-dimensional viewpoint the moduli
$T_{IJ}$ and $Y_{I}^{k}$  correspond to  the vev's of massless scalar
fields, members of the $N=4$  vector supermultiplets.
The explicit form of the $N=4$ heterotic partition function
\cite{nsw},\cite{kr}
$\Gamma^{SO(32)}_{(6,22)}[T_{IJ},Y_{I}^{k}]$ is:
\ba
\Gamma^{SO(32)}_{(6,22)}(T,Y)(\tau,{\bar \tau})&=&{\im }^{-3}({\rm det}
G_{IJ})^{3}\times\sum_{m^I,n^I}~
{\rm exp} \left[-\pi~T_{IJ}\frac{ (m^I+\tau n^I)(m^J+{\bar \tau}n^J)}
{\im } \right]\times \cr  
&&\frac{1}{2}\sum_{\gamma,\delta}
\prod_{k=1}^{16}{\rm exp}\left[~i\pi
\frac{1}{4}(n^I~Y^k_I Y^k_J~m^J+2 ~\delta~Y^k_I~n^I)\right]
\times~{\bar\th}\left[^{\gamma
+n^I~Y_I^k}_{\delta+m^I~Y_I^k}\right]({\bar
\tau})\cr 
&&\label{b3}
\ea
When all $Y$-moduli are zero $Y_{I}^{k}=0$ then the gauge group is
extended from
$G=U(1)^{22}$ to   $G=U(1)^6\times SO(32)$.

An alternative representation of $\Gamma_{(6,22)}$
is the one in which for $Y_{I}^{k}=0$ the extended gauge symmetry is
$G=U(1)^6\times E_8 \times E_8$ instead of $U(1)^6 \times SO(32)$:
\ba
\Gamma^{E_8 \times E_8}_{(6,22)}(T,Y)(\tau,{\bar \tau}) &=&{\im
}^{-3}({\rm det}
G_{IJ})^{3}~\times
\sum_{m^I,n^I}~
{\rm exp}\left[-\pi~T_{IJ}\frac{ (m^I+\tau n^I)(m^J+{\bar \tau}n^J)
}{\im }\right]\cr
\times \frac{1}{2}\sum_{\gamma_1,\delta_1}\prod_{k=1}^{8}
&&{\rm exp}\left[~i\pi
\frac{1}{4}(n^I~Y^k_I~Y^k_J~m^J+2
\delta_1~Y^k_I~n^I)\right]
\times{\bar\th}\left[^{\gamma_1
+n^I~Y_I^k}_{\delta_1+m^I~Y_I^k}\right]
({\bar\tau})\cr
\,\,\,\,\,\times \frac{1}{2}\sum_{\gamma_2,\delta_2}\prod_{k=9}^{16}~
&&{\rm exp}\left[~i\pi
\frac{1}{4}(n^I~Y^k_I~Y^k_J~m^J+2 \delta_2~Y^k_I~n^I)\right]
\times{\bar\th}\left[^{\gamma_2
+n^I~Y_I^k}_{\delta_2+m^I~Y_I^k}\right]
({\bar \tau})
\label{b4}\ea
Both the $SO(32)$ and $E_8\times E_8$ representations are connected
continuously by marginal deformations with  the $G=SO(44)$ maximal
gauge symmetry  point \cite{abk,klt}:
\def\wb{\bar W}
\be
\Gamma^{SO(44)}_{(6,22)}(\tau,{\bar \tau})=
{}~\frac{1}{2}~\sum_{\gamma,\delta}~\th^6
\left[^{\gamma}_{\delta}\right](\tau)
{\bar \th}^{22}\left[^{\gamma }_{\delta}\right]({\bar \tau}).
\label{b5}\ee

Another useful representation of the $\Gamma_{(6,22)}$ is that of the
lorenzian
left- and right-momentum even self dual lattice. This representation
is obtained by performing Poisson resummation on $m^I$, using either
$\Gamma^{SO(32)}_{(6,22)}(T,Y)$ or $\Gamma^{E_8 \times
E_8)}_{(6,22)}(T,Y)$:
\be
\Gamma_{(6,22)}~(P_I,{\bar P}_I, Q^k)~=~ \sum_{m_I, n^I,Q^k}
{\rm \exp}~i\pi\left[{\tau\over 2}P_I~g^{IJ}P_J-
{{\bar \tau}\over 2}{\bar  P}_Ig^{IJ}{\bar P}_J
-{\bar \tau }{\hat Q}^k{\hat Q}^k\right]
\label{b6}\ee
with
\be
{1\over 2}~P_I~g^{IJ}~P_J~-~{1\over 2}~{\bar  P}_I~g^{IJ}~{\bar
P}_J~-~{\hat Q}^k{\hat Q}^k~=
2~m_I~n^I~-~Q^k~Q^k~=~{\rm even~~integer}.
\label{b7}\ee
In the above equations $g^{IJ}$ is the inverse of $G_{IJ}$; the
lattice
momenta $P_I,~{\bar P}_I$, and the left--charges ${\hat Q}^k$ are
qiven in terms
the moduli parameters $G_{IJ}, B_{IJ}$ and $Y^k_I$ and in terms of
the charges
($m_I,~n^I,~Q^k$):
$$
P_I=
m_I+Y_I^kQ^k+{1\over2}Y_I^kY^k_J~n^J+B_{IJ}n^J+G_{IJ}n^J
$$
$$
{\bar P}_I=
m_I+Y_I^k~Q^k+{1\over2}Y_I^kY^k_J~n^J+B_{IJ}n^J-G_{IJ}~n^J
$$
\be
{\hat Q^k}~=~Q^k~+~Y_I^k~n^J.
\label{a1}\ee

All $N=4$ heterotic solutions are defined in terms of the vev's of
the moduli fields $(T_{IJ},~Y_I^k)$ and thus different solutions are
connected to each other by a string-Higgs phenomenon. The heterotic
$N=4$ spectrum is invariant under the target-space duality group
$SO(6,22;Z)$, e.g the generalization of the $R\rightarrow 1/R$
spectrum symmetry in a compactification on $S^1$ \cite{duality}. At
some special points of the moduli space we have extensions of the
gauge group as in the effective $N=4$ supergravity theories. In
string theories further extensions can take place, since due to the
non-zero winding charges ($n^I$) can become massless in special
points of the moduli space. Thus, in string theory, a large class of
disconnected $N=4$ supergravities are continuously related among
themselves due to the existence of the winding states. This precise
fact is the origin of the
perturbative string unification of all interactions in string
theories.

\section{$N=4 \rightarrow N=2$ spontaneous SUSY breaking}
\setcounter{equation}{0}

One of the defining characteristics of the $N=4$ theories is that the
states are classified by their transformation properties under the
$R$-symmetry group which, for $N=4$ supersymmetry is $G_R=SU(4)\sim
SO(6)$.
In the gravitational multiplet the gravitinos are in ${\bf 4}$
representation of $G_R$, the graviphotons are in ${\bf 6}$,  while
the graviton, the dilaton and the antisymmetric tensor field  are
singlets. The degrees of freedom of a massless $N=4$ vector multiplet
are also in definite representations of $G_R$: the scalars are in
${\bf 6}$, the gauginos
are in ${\bf 4}$, while the gauge bosons are singlets. In the
heterotic string $G_R$
is constructed in terms of the six--left moving compactified
supercoordinates, ($\Phi^I,~\Psi^I$). The world-sheet fermion
bilinears $\Psi^I~\Psi^J$ form an $SO(6)_{k=1}$ Kac--Moody algebra.
In the light-cone picture, the full spectrum of the theory is
classified in representations of $SO(6)_{k=1}$ and in terms of the
$U(1)_{0}$ helicity charge $q^0=\oint j^0$,
$~j^0=\Psi^{\mu}\Psi^{\nu}$, $\mu,\nu=3,4$. In the $N=4$ spectrum the
three internal helicity charges $q^i=\oint j^i$,
$~j^i=\Psi^{2k-1}\Psi^{2k}$, $k=1,2,3$ and $q^0$ are all
simultaneously integers for space-time bosonic states and
simultaneously half-integers for the fermionic states:
\vskip .2cm
$q^i=$ half~integers for space-time fermions
\vskip .2cm
$q^i=$ integers for space-time bosons.
\be ~ \label{bb6}\ee
 Furthermore all physical states have odd total $q^i$ charge
(GSO-projection)
\be
q^0~+~q^1~+q^2~+~q^3~=~{\rm odd~~integer}.
\label{bb7}\ee
The last condition remains valid for supersymmetric solutions with
less than four supersymmetries. In order to have a lower number of
supersymmetries, the $q^i$'s
must not be simultaneously integers or half-integers. It is then
necessary to
modify the world-sheet action $S^{2d}$, adding background fields that
can change
the individual values of the $q^i$'s, keeping however their total
$q^{i}$ charge:
\be
\Delta S^{2d}=\int dzd{\bar z}~F_{IJ}^a~(\Psi^I~\Psi^J~-~\Phi^I
\lrp ~\Phi^J)~{\bar J}^a,
\label{b8}\ee
where ${\bar J}^a$ denotes any dimension (0,1) operator. The part of
the left--moving operator $(\Phi^I\lrp \Phi^J)$ is necessary to
ensure the $N=(1,0)$ super-reparametrization  of the 2-d action. From
a higher--dimensional point of view, the $F^a_{IJ}$ denote
non-trivial gauge or gravitational (${\cal R}^{(KL)}_{IJ}$) field
backgrounds. In four dimensions they give rise to  non-vanishing
auxiliary fields. The permitted values of $F^a_{IJ}$ (${\cal
R}^{(KL)}_{IJ}$) are not arbitrary. Only those
for which
\be
U_L(F)=\exp~[\int dz~
F_{IJ}^a~(\Psi^I~\Psi^J~-~\Phi^I\lrp~\Phi^J)]
\label{b9}\ee
commutes with the 2-d super-current (~$T_F=\Psi^{\mu}~\partial
\Phi^{\mu}+\Psi^I~\partial \Phi^I$) are allowed. This restriction
generates   a quantization of the permitted $F^a_{IJ}$ (${\cal
R}^{(KL)}_{IJ}$) backgrounds.

A partial $N=4~\rightarrow ~N=2$ breaking is possible \cite{ss1} when
$F^a_{3,4}=-F^a_{5,6}= H$ is not zero (self-duality condition).
Indeed, in that case the $q^2$ and $q^3$ charges are shifted,
preserving the total $q^i$ charge. In order to define the full
deformation of the spectrum it is necessary to find a representation
of the partition function in which  the bosonic charges
\be
Q^B_2=\oint dz \Phi^3\lrp \Phi^4
{}~{\rm and}~
Q^B_3=\oint dz \Phi^5\lrp \Phi^6
\label{b10}\ee
are well defined. As a starting point we fermionize the four internal
bosonic coordinates
\be
\partial \Phi^I=y^Iw^I, ~{\rm and}~ {\bar \partial} {\bar
\Phi}^I={\bar y}^I~{\bar w}^I, ~I=3,4,5,6.
\label{b11}\ee
In this representation the 2d supercurrent is \cite{abk},\
\be
T_F=\Psi^{\mu}~\partial \Phi^{\mu}~+\sum_{I=1}^2 \Psi^I~\partial
\Phi^I~+\sum_{I=3}^6 \Psi^I~y^I~w^I.
\label{b12}\ee
We will now perform the following $Z_4$ transformation:
$$
\Psi^3 \rightarrow ~~\Psi^4, ~~~ y^3 \rightarrow ~~y^4,~~~\Psi^5
\rightarrow -\Psi^6, ~~~ y^5 \rightarrow -y^6,
$$
$$
\Psi^4 \rightarrow -\Psi^3, ~~~ y^4 \rightarrow -y^3,~~~~~\Psi^6
\rightarrow ~~\Psi^5, ~~~ y^6\rightarrow ~~y^5,
$$
$$
w^3\rightarrow ~w^4,~~~~w^4\rightarrow ~w^3,~~~~~~w^5\rightarrow
{}~w^6,~~~~~~w^6\rightarrow ~w^5,
$$
\be ~~ \label{b13}\ee
which leaves (\ref{b12}) invariant.
The  above transformation corresponds to a $\pi /2$ rotation on the
complex fermion basis:
$$
\chi_1 \rightarrow ~ e^{2i\pi\phi}~\chi_1~~~{\rm
with}~~~\chi_1~=~{{\Psi^3~+~i\Psi^4}\over
\sqrt{2}}~~~~~~~~~~~
$$
$$
\chi_2 \rightarrow ~ e^{-2i\pi\phi}~\chi_2~~~{\rm
with}~~~\chi_2~=~{{\Psi^5~+~i\Psi^6}\over
\sqrt{2}}~~~~~~~~~~~
$$
$$
Y_1 \rightarrow ~ e^{2i\pi\phi}~Y_1~~~{\rm
with}~~~Y_1~=~{{y^3~+~iy^4}\over \sqrt{2}}~~~~~~~~~~
$$
$$
Y_2 \rightarrow ~ e^{-2i\pi\phi}~Y_2~~~{\rm
with}~~~Y_2~=~{{y^5~+~iy^6}\over \sqrt{2}}~~~~~~~~~~
$$
$$
W_{-} \rightarrow  e^{4i\pi \phi}W_{-}~{\rm
with}~W_{-}={{(w^3-w^4)+i(w^5-w^6)}\over {2}}
$$
$$
W_{+} \rightarrow W_{+}~~~{\rm
with}~~~W_{+}={{(w^3+w^4)+i(w^5+w^6)}\over {2}}
$$
\be ~~~\label{b14}\ee
Similarly for the right--moving degrees of freedom (${\bar \Psi}^I$,
${\bar y}^I, ~{\bar w}^I,~I=3,4,5,6$).
The above transformation is a symmetry only if the rotation angle is
a multiple of $\pi/2$ or  $\phi~=k/4$, with $k$ integer.

Observe that with the help of the word sheet fermions we  can
classify the $N=4$ string spectrum  in terms  of a left and right
$U(1)$ charges
$Q_L=\oint j_L$ and $Q_R=\oint j_R$, where
\ba
j_L&=&\chi_1\chi^{\dagger}_1-\chi_2\chi^{\dagger}_2+
Y_1Y^{\dagger}_1-Y_2Y^{\dagger}_2
+2W_{-}W^{\dagger}_{-}~,\cr
Q_L&=&q_{\chi_1}-q_{\chi_2} + q_{Y_1}-q_{Y_2}+2q_{W_-}
\label{b15}\ea
and
$$
j_R={\bar \chi}_1{\bar \chi}^{\dagger}_1-
{\bar \chi}_2{\bar
\chi}^{\dagger}_2+{\bar Y}_1{\bar Y}^{\dagger}_1-
{\bar Y}_2 {\bar Y}^{\dagger}_2
+2{\bar W}_{-} {\bar W}^{\dagger}_{-},~~~~~~~~~
$$
\be
Q_R=q_{{\bar
\chi}_1}-q_{{\bar \chi}_2}+{\bar q}_{Y_1}-
{\bar q}_{ Y_2}+2{\bar
q}_{W_-}
\label{b16}\ee

We are now in a position to switch on non-vanishing $F_{IJ}^a$ by
performing a boost among the fermionic charge lattice  and  the
$\Gamma_{(2,n)}$ lattice:
$$
q_{\chi_1}~\rightarrow~q_{\chi_1}~+~h_in^i,
{}~~~~~~~q_{\chi_2}~
\rightarrow~ q_{\chi_2}~-~h_i n^i
$$
$$
q_{{\bar \chi}_1} ~\rightarrow~q_{{\bar
\chi}_1}~+~h_in^i,~~~~~~~
q_{{\bar \chi}_2}~\rightarrow~ q_{{\bar
\chi}_2}~-~h_i n^i
$$
$$
q_{Y_1}~\rightarrow~q_{Y_1}~+~h_in^i,
{}~~~~~~~q_{Y_2}~\rightarrow~q_{Y_2
}~-~h_in^i
$$
$$
q_{{\bar Y}_1}~\rightarrow~q_{{\bar Y}_1}~+~h_in^i,
{}~~~~~~~q_{{\bar
Y}_2}~ \rightarrow~q_{{\bar Y}_2}~-~h_in^i
$$
$$
q_{W_-}~\rightarrow~q_{W_-}~+~2h_in^i,
{}~~~~~q_{W_+}~\rightarrow~q_{W_+}
{}~~~~~~~~
$$
$$
q_{{\bar W}_-}~\rightarrow~q_{{\bar W}_-}~+~2h_in^i,
{}~~~~~q_{{\bar
W}_+}\rightarrow~q_{{\bar W}_+}~~~~~~~~
$$
$$
P^L_i(h_i)~=~P^L_i~-~h_i~(Q_L-Q_R,), 
$$
\be
P^R_i(h_i)~=~P^R_i~-~h_i~(Q_L
-Q_R)
\label{b17}\ee
with
$$
P^L_i=m_i~+~Y_i^a~Q^a~+~{1\over2}Y_i^a~Y^a_j~n^j~
+B_{ij}~n^j~+~G_{ij}~
n^j
$$
$$
P^R_i=m_i~+~Y_i^a~Q^a~+~{1\over2}Y_i^a~Y^a_j~n^j~
+B_{ij}~n^j~-~G_{ij}~
n^j
$$
\be ~~~~\ee
$Y^a_i$ $i=i,2$, $a=1,2,...,18$ are the Wilson-line moduli of the
$\Gamma (2,18)$ lattice.
Owing to the non-zero $h_i$ shift, two of the $N=4$ gravitinos
become massive, with mass proportional to $|q_{\chi_1}-q_{\chi_2}|$.
The $N=4$ gravitinos
have vanishing $m_i, n^i, q_{Y_i}, q_{{\bar Y}_i}, q_{W_{i}},
q_{{\bar W}_{i}}$ charges. The two of them remain massless since
$|q_{\chi_1}-q_{\chi_2}|=0$, while for the other two   become massive
since $|q_{\chi_1}-q_{\chi_2}|=1$:
\be
(m^2_{3/2})_{1,2}=0, ~~~(m^2_{3/2})_{3,4}={|F|^2\over 4{\rm
Im}T{\rm Im}U}, ~~~
\label{b18}\ee
${\rm with}~F~=~h_1~+U~h_2$, $T$ and $U$ are usual complex moduli of
the $\Gamma_{(2,2)}$ lattice:
$$
T=i\sqrt{{\rm det} G_{ij}}~+~B_{12}, ~~~~~~~~~~~~~~~~~~~~~~~~~
{}~~~~~~~~~~
$$
\be
U = {(i\sqrt{{\rm det}
G_{ij}}~+~G_{12})\over G_{22}}.
\label{tu}\ee

The global existence of the supercurrent implies in this case the
quantization condition:
$4h_i=$ integer. The $N=2$ partition function $\Z^{4\rightarrow
2}(F)$ is obtained from that of $N=4$ by shifting  the lattice
momenta $P_i$ and the $R$--charges $q_i$ as above. Performing a
Poisson resummation on $m_i$, we obtain the following expression:
\vskip .2cm
$$
\centerline{ $ \gamma = 2h_i n^i.~~~\delta = 2h_i m^i,~~~F=h_1+U~h_2$
}
$$
\ba
\Z^{4\rightarrow 2}(F)&=& {(\im)^{-1} \over \eta^2{\bar
\eta}^2}\sum_{m^i,n^i}~{1\over 2}~
\sum_{\alpha, \beta}~ (-)^{\alpha+\beta +\alpha \beta}~
\times \frac{\th^{2}[^{\alpha}_{\beta}]}{\eta^2}\frac{\th[^{\alpha~+~
\gamma}_{\beta~+~\delta}]}{\eta}
\frac{\th[^{\alpha~-~\gamma}_{\beta~-~\delta}]}{\eta}
{}~{\Gamma_{(2,2)}[^{n^i}_{m^i}]\over
|\eta|^4}~Z_{(4,4)}[^{\gamma}_{\delta}]\cr
&\times&\sum_{{\bar \alpha}, {\bar \beta}}~
\frac{1}{2}~\sum_{\bar \alpha, \bar \beta}
\frac{{\bar \th}^{6}[^{\bar \alpha}_{\bar \beta}]}{{\bar \eta}^6}
{}~\frac{{\bar \th}[^{{\bar \alpha}~+~{\bar \gamma}}_{{\bar
\beta}~+~{\bar \delta}}]}{{\bar \eta}}~\frac{{\bar \th}[^{{\bar
\alpha}~-~{\bar \gamma}}_{{\bar \beta}~-~{\bar \delta}}]}{{\bar
\eta}}
{}~\sum_{\epsilon, \zeta}\frac{1}{2}~\frac{{\bar
\th}^{8}[^{\epsilon}_{\zeta}]}{{\bar \eta}^8},
\label{b19}\ea
where
\be
\Gamma_{(2,2)}[^{n^i}_{m^i}]~=~{\sqrt {{\rm det}~G_{ij}}}~(\im)^{-1}
\times 
{\rm exp}~[-\pi G_{ij} \frac{(m^i+n^i\tau)(m^j+n^j~{\bar
\tau})}{\im}+2i \pi B_{ij}m^in^j]
\ee
and
\be
{}~Z_{(4,4)}[^{\gamma}_{\delta}]~=
\frac{1}{2}\sum_{a,b} \frac{|\th[^{a}_{b}]|^{2}}{|\eta|^{2}}
{}~\frac{|\th[^{a~+~2\gamma}_{b~+~2\delta}]|^{2}}{|\eta|^{2}}
{}~\frac{|\th[^{a~+~\gamma}_{b~+~\delta}]|^{2}}{|\eta|^{2}}
{}~\frac{|\th[^{a~-~\gamma}_{b~-~\delta}]|^{2}}{|\eta|^{2}}.
\label{b20}\ee
When $h_i=0$  $(\gamma,~\delta=0)$,  $\Z^{4\rightarrow 2}(F=0)$
corresponds to
the $N=4$ heterotic string solution based on a gauge group
$U(1)\times U(1) \times SO(8) \times E_8 \times E_8$;  the $SO(8)$
gauge group factor corresponds to the extended symmetry of the
$\Gamma_{(4,4)}$ lattice at the fermionic point
\be
{}~Z_{(4,4)}[^0_0]~=~\frac{1}{2}~\sum_{a,b}~
{}~\frac{|\th[^{a}_{b}]|^{8}}{|\eta|^{8}}.
\label{b21}\ee
The sum over $m^i$ and $n^i$ gives rise to the $\Gamma (2,2)$ lattice
at an arbitrary point of the moduli space:
\be
\sum_{m^i,n^i}~\Gamma_{(2,2)}[^{n^i}_{m^i}]~=
{}~\Gamma_{(2,2)}[T,U].
\label{b22}\ee
When $h_i~\ne~0$ $(\gamma,~\delta)=(2h_in^i,~2h_im^i)$,  then the
$N=4$ supersymmety is spontaneously broken to $N=2$ and the gauge
group is reduced to $U(1)^2 \times E_7 \times E_8$, as in orbifold
models.

The important difference between the $N=2$ model described above and
the orbifold models \cite{orbif} of order {\bf N} is in the
parameters $\gamma$ and $\delta$, which appear as arguments in
$\th$-functions. In the model in which somme of the $N=4$ the
supersymmetries are broken spontaneously, $\gamma=2h_in^i$ and
$\delta=2h_im^i$ are not independent but are given in terms of the
$h_i$ and in terms of the charges $n^i,~m^i$ of the $\Gamma
(2,2)[^{n^i}_{m^i}]$ lattice.
In the standard symmetric  orbifolds of order {\bf N}, the arguments
$\gamma$ and $\delta$, ($\gamma~=~2l$/{\bf N} and  $\delta~=~2k$/{\bf
N} with  $l,k$ = 0,1,...,{\bf N} -1 ), are independent  arguments;
their summation gives rise to the orbifold projections and to some
additional states in the twisted sector:
\ba
\Z^{N=2}_{\rm orb}&=&{(\im)^{-1} \over {|\eta(\tau
|^4}}~~\frac{1}{\bf N}~\sum_{\gamma,~\delta=0}^{ 1}
{}~\sum_{\alpha,~\beta}~(-)^{\alpha+\beta +\alpha
\beta}\cr 
&\times& \frac{\th^{2}[^{\alpha}_{\beta}](\tau)}{\eta^2(\tau)}
\frac{\th[^{\alpha~+~\gamma}_{\beta~+~\delta}](\tau)}{\eta(\tau)}
\frac{\th[^{\alpha~-~\gamma}_{\beta~-~\delta}](\tau)}{\eta(\tau)}
{\Gamma_{(2,2)}\over
{|\eta(\tau)|^4}}~Z_{(4,4)}[^{\gamma}_{\delta}]\cr
&\times&\frac{1}{2}~\sum_{\bar \alpha, \bar \beta}
\frac{{\bar \th}^{6}[^{\bar \alpha}_{\bar \beta}]({\bar \tau})}{{\bar
\eta}^6({\bar \tau})}
{}~\frac{{\bar \th}[^{{\bar \alpha}~+~{\bar \gamma}}_{{\bar
\beta}~+~{\bar \delta}}]({\bar \tau})}{{\bar \eta}({\bar
\tau})}~\frac{{\bar \th}[^{{\bar \alpha}~-~{\bar \gamma}}_{{\bar
\beta}~-~{\bar \delta}}]({\bar \tau})}{{\bar \eta}({\bar
\tau})}
\times \frac{1}{2}\sum_{\epsilon, ~\zeta}~\frac{{\bar
\th}^{8}[^{\epsilon}_{\zeta}]({\bar \tau})}{{\bar \eta}^8({\bar
\tau})}.
\label{b23}\ea
In the language of orbifolds, the spontaneously broken theory,
$\Z^{4\rightarrow 2}$, corresponds to a {\em freely acting orbifold}.
Indeed, using the quantization condition,
\be
{\rm \bf N}~h_i~=~{\rm integer}
\label{b24}\ee
and the mod~$2$ periodicity properties of the $\th$-functions in the
arguments
\be
\th[^{a+2k}_{b+2l}]~=~\th[^{a}_{b}]~e^{i\pi la},
\label{b25}
\ee
it is possible  to write the $\Z^{4\rightarrow 2}$ theory in orbifold
language.
In order to make this correspondence explicit, we must first redefine
the lattice charges $n^i={\rm \bf N}{\hat n}^i+\gamma^i $ and
$m^i={\rm \bf N}{\hat m}^i+\delta^i$. Thanks to the property
(\ref{b25}),  the above lattice charge redefinition makes the
arguments of the $\th$-functions independent of ${\hat n}^i$ and
${\hat m}^i$; they depend only on ${\hat \gamma}~=~2h_i\gamma^i$ and
${\hat \delta}~=~2h_i\delta^i$. Performing now a Poisson resummation
on ${\hat m}^i$, we  obtain the orbifold representation for
$\Z^{4\rightarrow 2}$ theory:
\ba
\Z^{4\rightarrow 2}(F)&=& {(\im)^{-1} \over \eta^2(\tau){\bar
\eta}^2({\bar \tau})}{1\over {\bf \rm
N}}\sum_{\gamma^i~\delta^i}~{1\over 2}~\sum_{\alpha,
\beta}~(-)^{\alpha+\beta  +\alpha \beta}
\times
\frac{\th^{2}[^{\alpha}_{\beta}](\tau)}{\eta^2(\tau)}\frac{\th[
^{\alpha~+~{\hat \gamma}}_{\beta~+~{\hat \delta}}](\tau)}{\eta(\tau)}
\frac{\th[^{\alpha~-~{\hat \gamma}}_{\beta~-~{\hat
\delta}}](\tau)}{\eta(\tau)} \cr
&\times& \sum_{{\bar \alpha}, {\bar \beta}}~
\frac{1}{2}~\sum_{\bar \alpha, \bar \beta}
\frac{{\bar \th}^{6}[^{\bar \alpha}_{\bar \beta}]({\bar \tau})}{{\bar
\eta}^6({\bar \tau})}
{}~\frac{{\bar \th}[^{{\bar \alpha}~+~{\hat \gamma}}_{{\bar
\beta}~+~{\hat \delta}}]({\bar \tau})}{{\bar \eta}({\bar
\tau})}~\frac{{\bar \th}[^{{\bar \alpha}~-~{\hat \gamma}}_{{\bar
\beta}~-~{\hat \delta}}]({\bar \tau})}{{\bar \eta}({\bar
\tau})}\cr
&\times& Z^{SO(8)}_{(4,4)}[^{\hat \gamma}_{\hat
\delta}]~~\frac{\Gamma_{(2,2)}[^{\gamma^i}_{\delta^i}]}{|\eta
(\tau)|^4}
{}~\sum_{\epsilon, \zeta}\frac{1}{2}~\frac{{\bar
\th}^{8}[^{\epsilon}_{\zeta}]({\bar \tau})}{{\bar \eta}^8({\bar
\tau})},
\label{b26}\ea
where
\be
\Gamma_{(2,2)}~[^{~\gamma^i}_{~\delta^i}]~= 
\sum\exp i\pi \left[
{2\delta^i{\hat m}_i \over {\bf \rm N}}+ \tau {1\over 2}
P^L_ig^{ij}P^L_j - {\bar \tau}{1\over 2} P^R_ig^{ij}P^R_j \right],
\label{b27}\ee
with
\be
 P^L_i~=~{\hat m}_i~+~\left({\hat n}^j+{\gamma^j \over {\bf \rm
N}}\right)G_{ij}~~{\rm and}~~
 P^R_i~=~{\hat m}_i~-~\left({\hat n}^j+{\gamma^j \over {\bf \rm
N}}\right)G_{ij}.
\label{b28}\ee
The connection of $\Z^{4\rightarrow 2}$ with the freely acting
orbifolds gives us the way to switch all the  moduli of $Z_{(4,4)}$
and so to move out of the  $SO(8)$ extended symmetry point. This
extension can be done by  replacing the $SO(8)$ characters of
$Z^{SO(8)}_{(4,4)}[^{\gamma}_{\delta}]$,  which are defined  at the
fermionic point
\be
Z_{(4,4)}^{SO(8)}[^{\gamma}_{\delta}]~=~
\frac{1}{2}\sum_{a,b}
\frac{|\th[^{a}_{b}]|^{2}}{|\eta(\tau)|^{2}}
\frac{|\th[^{a+2\gamma}_{b+2\delta}]|^{2}}
{|\eta(\tau)|^{2}}
\frac{|\th[^{a+\gamma}_{b+\delta}]|^{2}}
{|\eta(\tau)|^{2}}
\frac{|\th[^{a-\gamma}_{b-\delta}]|^{2}}
{|\eta(\tau)|^{2}} ~~~~~~~~~~~~~~~~~~~
\label{b29}\ee
by the characters of the $Z_{\bf \rm N}$-orbifold,
$Z^{(twist)}_{(4,4)}[^{\gamma}_{\delta}]$.

$\bullet$ When $(\gamma,~\delta) = (0,0)$,
$Z_{(4,4)}^{SO(8)}[^{0}_{0}]$ must be replaced by the ``untwisted"
orbifold  partition function, which depends on the $T_{IJ}$ moduli
\be
Z_{(4,4)}^{SO(8)}[^{0}_{0}] \rightarrow
Z_{(4,4)}[^{0}_{0}][T_{IJ}]={\Gamma_{(4,4)}[T_{IJ}] \over
|\eta(\tau)|^8},
\ee

$\bullet$ When $(\gamma,~\delta)~\ne~(0,0)$,
$Z_{(4,4)}^{SO(8)}[^{\gamma}_{\delta}]$ {\it no modification is
necessary} since  the ``twisted" orbifold  partition function remains
the same at any point of the moduli space:
\be
Z_{(4,4)}^{SO(8)}[^{\gamma}_{\delta}]\rightarrow
Z^{twist}_{(4,4)}[^{\gamma}_{\delta}]~=~Z_{(4,4)}^{(SO(8)}~[^{\gamma}_ 
{\delta}]~~~~ (\gamma,~\delta)~\ne~(0,0).
\label{b30}\ee
The models described above are special cases of a general class of
models
having the interpretation of freely acting orbifods of the $N=4$
heterotic string theory.
They are obtained in the following way.
Consider $\Gamma(6,22)$  and set the appropriate moduli to
special values, so that it factorizes as
\be
\Gamma_{(6,22)}\to \Gamma_{(2,18)}\;\Gamma_{(4,4)}
\label{b32}\ee
Now consider the orbifold that acts as a $Z_{\bf \rm N}$ rotation on
$\Gamma_{(4,4)}$
and as a translation by an ${\bf \rm N}$-th lattice vector
$\vec\varepsilon/{\bf \rm N}$ with
$\vec\varepsilon=(\vec\varepsilon_L;\vec\varepsilon_R,\vec \zeta)$,
on $\Gamma_{(2,18)}$.
$\vec\varepsilon_{L,R}$ are two--dimensional vectors while
$\vec\zeta$ is a sixteen dimensional vector.
Owing to the accompanying translation on $\Gamma_{(2,18)}$, this is a
freely acting orbifold.

The two types of constructions of $N=4\to N=2$ theories we have
presented above, have complementary features.
In the first approach of using a specific generalized boost at the
fermionic point, it is evident that there is a {\it one-to-one
correspondence} of states
between the original $N=4$ supersymmetric theory and the final {\it
spontaneously
broken} $N=4\to N=2$ theory. This is what should be expected during
spontaneous
symmetry breaking.
In the second, {\it freely acting orbifold} approach, we have a clear
geometrical intuition about the spontaneously broken theory, which
will be very useful for the  identification of the heterotic and type
II dual theories.

Inspection of the standard $N=4$ gravitino vertex operators shows
that two of them are invariant while the other two transform, one
with a phase $e^{2\pi i/{\bf \rm N}}$ and the other with $e^{-2\pi
i/{\bf \rm N}}$.
In order for them to survive in the spectrum they have to pair up
with a state
of the (2,18) lattice carrying momentum $p=(\vec m;\vec n,\vec Q)$
but no oscillators (these will shift the mass to the Planck scale).
Since such a lattice state picks up a phase 
$e^{2\pi i \e \cdot p/{\bf \rm N}}$,
one of the two massive gravitinos will have momentum $p_1$ with the
property
that $p_1\cdot \e=1~~$mod$~{\bf \rm N}$ while the other $p_2$ with
$p_2\cdot \e=-1$~~mod~{\bf N}.
The mass formulae given in (\ref {b18}) are special cases of the
above.

There is an essential difference betweenthe models with spontaneous
breaking of the $N=4\to N=2$ and the standard  $N=2$ orbifold models.

$\bullet$ First, in the spontaneously broken case, one expects an
effective
restoration of the $N=4$ supersymmetry in a corner of the moduli
space $T,~U$, where the two massive gravitinos become light,
$m_{3/2}\to 0$.

$\bullet$ Second, in the standard obifolds there is no restoration of
the $N=4$
supersymmetry at any point of the moduli space.

 If there is an effective restoration of the $N=4$ supersymmetry in
the spontaneously broken case, then one must find zero higher-genus
corrections to the coupling constants of the theory in the $N=4$
restoration limit $m_{3/2}\to 0$. This restoration phenomenon has
been checked in ref.\cite{kkpr1,kkpr2} where the  one-loop
corrections of the coupling constants were performed  for a class of
$Z_2$ models based on $E_8\times E_7\times SU(2)\times U(1)^2$ gauge
group. A more detailed discussion of the general heterotic models and
their type II duals will appear in ref. \cite{kkpr3}.
Here I will restrict myself to the case of $Z_2$ freely acting
orbifolds with $F~=h_1+Uh_2~=~1/2$. The  $m_{3/2}\to 0$ limit in this
class of models  corresponds  to the  corner of the moduli space
${\rm Im}T~{\rm Im}U \to \infty$, which implies an effective
decompactification of one  of the two coordinates  of
$\Gamma_{2,2}(T,U)$,  ($R_1\to \infty$ and $R_2$ arbitrary; ${\rm
Im}T \sim R_1 R_2,~~{\rm Im}U \sim R_1/R_2$). In this limit, $T,~U\to
\infty$, one expects vanishing corrections to the coupling constants
due to the effective  $N=4$  restoration. Using the explicit results
of ref. \cite{kkpr1},
\be
\Delta^{freely}_{(8,7)}~=~{16 \pi^2 \over g^2_{E_8}}~-{16 \pi^2 \over
g^2_{E_7}}~= 
\delta b~~{\rm log}~\left[|\mu|^2~{\rm Im}~ T{\rm Im}
{}~U|\th_4(T)~\th_4(U)|^4\right],
\label{b33}\ee
where $\delta b=b_8-b_7$ and $b_i$ are the $\beta$-function
coefficients due to massless particles.  When $T$ and $U$  are large,
${\rm Im}T~{\rm Im}U~\gg~1$, due to the asymptotic behaviour of
$\th_4(T)=1+{\rm {\cal O}}(e^{-{i\pi~T}})$:
\be
\Delta^{freely}_{(8,7)}~~\rightarrow ~~\delta b~{\rm
log}~|\mu|^2~{\rm Im} T~{\rm Im} U~.
\ee
The logarithmic contribution is an  artefact due to the infrared
divergences.
In fact by turning on Wilson lines appropriately (e.g. small Higgs
vev's of the vector multiplets), we can arrange that there are no
charged states with masses $\mu^2_W ~\sim ~|W|^2/{\rm Im} T~{\rm Im}
U~$ below $m^2_{3/2}$. In this case the logarithmic term becomes:
\be
\delta b~{\rm log}~|\mu|^2~{\rm Im}T~{\rm Im} U ~~\rightarrow
{}~~\delta
b~{\rm log}~{\mu^2_W \over m^2_{3/2}+\mu^2_W  }
\sim~{\rm \cal O}
\left({m^2_{3/2} \over \mu^2_W } \right);
\ee
the logarithmic divergence thus disappears and the thresholds vanish,
which shows the restoration of the $N=4$ supersymmetry in the light
massive gravitino limit as expected.
 In the calculation of individual couplings, there is an extra
contribution $Y(T,U)$, which is ``universal" for $g_{E_8}$ and
$g_{E_7}$; the explicit calculation in ref.
\cite{kkpr1}--\cite{kkpr3} shows
that $Y(T,U)$ behaves  like
\be
Y(T,U ) \rightarrow {m^2_{3/2}\over M^2_s}\;\;\;{\rm as}\;\;\;
m_{3/2}\to 0.
\label{b48}\ee
Thus individual couplings also vanishe in the limit $m_{3/2}\to 0$.

In the standard orbifold with $N=2$ space-time supersymmetry, the
corrections to the coupling constants have a different behaviour for
$T,~U\gg 1$ \cite{dkl,kkpr2}:
\be
\Delta^{orb}_{(8,7)}=\delta b~\log\left[\mu^2~{\rm Im}T{\rm Im}
U|\eta(T)\eta(U)|^4 \right].
\label{b40}\ee
When $T,U$ is large, $ImT ImU \gg~1$,
\be
\Delta^{orb}_{(8,7)}~~\rightarrow ~~\delta b~\left[{\pi \over 3}({\rm
Im}T+{\rm Im}U)~+~ \log~{|W|^2 \over M^2_s} \right]
+{\rm
finite~terms}.
\label{b41}\ee
So, in the standard orbifolds, the correction to the coupling
constants grows linearly with the  five--dimensional volume. This
shows that the $N=2$ supersymmetry is ``not extended" in the
decompactification limit $R_1\to\infty$. On the other hand there is
an extension of the supersymmetry in the freely acting orbifold case.

In the opposite limit ${\rm Im}T {\rm Im}U\to 0$, the situation is
different:

 i) In  the freely acting orbifolds the two massive gravitinos
becomes superheavy: $m_{3/2}\to \infty$ in the limit ${\rm Im}T {\rm
Im}U\to 0$.

ii) In the  standard orbifolds, thanks to the duality symmetry
$R_i\to 1/R_i$ the behaviour $T,U\to 0$, is identical to the dual
model with  $T'=-1/T,~U'=-1/U~\to \infty$ and thus
\be
\Delta^{orb}_{(8,7)}(T,U,W)=\Delta^{orb}_{(8,7)}(T',U',W')\rightarrow
\delta b~\left[{\pi \over 3}({\rm Im}T'+{\rm Im}U')+\log{|W'|^2 \over
M^2_s}\right]+{\rm finite~terms}
\label{b41o}\ee
In the freely acting orbifolds, the $SO(2,2;Z)$ duality symmetry  is
reduced to a smaller subgroup due to the $Z_2$ action on the lattice.
Thus one expects non-restoration of the $N=4$ supersymmetry  in this
limit ( $T,~U\to 0$ $m_{3/2}\to\infty$):
$$
\Delta^{freely}_{(8,7)}~=~\delta b'~\left[~{\rm
log}|\th_2(T')~\th_2(U')|^4 +{\rm log}{|W'|^2 \over
M^2_s}\right].
$$
\be ~ \label{b41f}\ee
In the above equation  we have used the $\th$-identity
\be
{\rm Im}T|\th_4(T)|^4={\rm Im}T'|\th_2(T')|^4, ~~~T'=-{1 \over
T}.
\label{b41th}\ee
Using the asymptotic behaviour for $T',U'\gg~1$ of
${\rm
log}~|\th_2(T')~\th_2(U')|^4 $ one obtains:
\be
\Delta^{freely}_{(8,7)}\rightarrow \delta b'\left[{\pi \over
3}({\rm Im}~T'+{\rm Im}~U')+ \log{|W'|^2 \over M^2_s}\right]
+{\rm finite~terms}.
\label{b41as}\ee
It is interesting to observe that the $m_{3/2}\to\infty$ limit
\cite{kkpr1} of the freely acting orbifolds corresponds  to a corner
in the moduli space of $T,U$  where the two classes of theories (the
freely  and non-freely acting orbifolds)  ``touch" each other.
Both theories are effectively five-dimensional. Thus the
five-dimensional standard $N=2$ orbifolds can be viewed as an
$m_{3/2}\to \infty$ limit of some spontaneously broken $N=4$ models.

\section{$N=4 \rightarrow N=1$ spontaneous SUSY breaking}
\setcounter{equation}{0}

Using the connection between the freely acting orbifolds and the
spontaneous breaking $N=4\rightarrow N=2$, we can proceed to further
break  the supersymmetry to $N=1$.
We will restrict ourselves  to the  case where the possible quantized
 parameters are of order  {\bf N}=2 $2|h^i|=1$. In that case the
spontaneously broken $N=4\rightarrow N=1$ theory is strongly
connected to $Z^2\times Z^2 $ freely acting orbifolds; the $Z^2\times
Z^2$ acts simultaneously as a  rotation on the coordinates
$\Phi^I,~\Phi^J$ and $\Psi^I,~\Psi^J$ of the two complex planes and
as  a translation on the third complex plane $\Phi^L$. Denoting  by
$\Phi_A,~A=1,2,3$, the complex internal coordinates and by
$\chi_A,~A=1,2,3$, the three complex fermionic world-sheet
superpartners, the non-trivial actions of the orbifold are:
\vskip .2cm
1)~~~$\Phi_1 \rightarrow \Phi_1+2\pi h_1$,
$~~~(\Phi_2,~\chi_2)~\rightarrow e^{i2 \pi h_1}(\Phi_2,~\chi_2)$,
$~~~(\Phi_3,~\chi_3)~\rightarrow e^{-i2\i h_1}(\Phi_3,~\chi_3)$.
\vskip .2cm

2)~~~$\Phi_2 \rightarrow \Phi_2+2\pi h_2$,
$~~~(\Phi_1,~\chi_1)~\rightarrow e^{i2\pi h_2} (\Phi_1,~\chi_1)$,
$~~~(\Phi_3,~\chi_3)~\rightarrow e^{-i2\pi h_2}(\Phi_3,~\chi_3)$.

\vskip .2cm
3)~~~$\Phi_3 \rightarrow \Phi_3+2\pi h_3$,
$~~~(\Phi_1,~\chi_1)~\rightarrow e^{i2\pi h_3} (\Phi_1,~\chi_1)$,
$~~~(\Phi_2,~\chi_2)~ \rightarrow e^{-i2\pi h_3}(\Phi_2,~\chi_2)$.
\vskip .2cm
In order to obtain the partition function and define the theory, we
need to introduce
the ``shifted" and ``twisted" characters of the three complex
coordinates.
We denote by $(\gamma_A,~\delta_A)$ the translation shifts and by
$(H_A,~G_A)$ the rotation twists.
When the ``twist" is zero $(H_A,~G_A)=(0,~0)$:
\be
Z_A\left[^{\gamma_A
{}~;~0}_{\delta_A~;~0}\right]~= ~{\Gamma_{(2,2)}~
[^{~\gamma_A}_{~\delta
_A}] \over |\eta|^4}~\delta(H_A)\delta(G_A).
\ee
When the twist is non-zero $(H_A,~G_A)\ne(0,~0)$:
\be
Z_A\left[^{\gamma_A ~;~H_A}_{\delta_A~;~G_A}\right]~=~{1\over
2}Z^{twist}_{(2,2)}\left[^{H_A}_{G_A}\right]\times 
\left[~\delta (\gamma_A
)~\delta(\delta_A )~+~~\delta (\gamma_A +H_A)~\delta(\delta_A
+G_A)\right].
\ee
In the above equation, the
$Z^{twist}_{(2,2)}\left[^{H_A}_{G_A}\right]$ can be written either in
terms of a twisted boson or in terms of 2d-fermionic characters with
shifted boundary conditions:
\ba
Z^{twist}_{(2,2)}\left[^{H_A}_{G_A}\right]&=&{4|\eta|^2 \over
\th(^{1~+~H_A}_{1~+~G_A})\th(^{1~-~H_A}_{1~-~G_A})|}=
{1 \over 2}~\sum_{a,b}~{|\th^2(^a_b)
\th(^{a~+~H_A}_{b~+~G_A})\th(^{a~-~H_A}_{b~-~G_A})| \over |\eta|^4}\cr\cr
 &&{\rm
if }~~~(H_A,~G_A)\ne (0,0)
\ea
The world-sheet modular properties of
$Z_A\left[^{\gamma_A ~;~H_A}_{\delta_A~;~G_A}\right]$ are the same
for any point of the moduli space and thus at the $SO(4)_A$ fermionic
point, which takes the following form:
\be
Z_A\left[^{\gamma_A ~;~H_A}_{\delta_A~;~G_A}\right]|_{T^A_0,U^A_0}~=
{1 \over 2} \sum_{a,b}{|\th^2(^a_b)\th(^{a+H_A}_{b+G_A})
\th(^{a-H_A}_{b-G_A})| \over
|\eta|^4}e^{i\pi(a \delta^A+b \gamma^A+\gamma^A \delta^A)}.
\ee
The above expression makes  the world-sheet modular properties  of $
Z_A\left[^{\gamma_A ~;~H_A}_{\delta_A~;~G_A}\right]$ more transparent
under $SL(2,Z)_{\tau}$; it also  makes  clear the connection to the
fermionic models.
The role of the  phase factor
$e^{i\pi~(~a~\delta^A~+~b~\gamma^A~+~\gamma^A~\delta^A)}$ is of main
importance, since it clarifies the way we had to choose the
coefficient of the fermionic characters.

We are now in a position to construct consistent $N=4\to N=1$ models
using the fermionic construction algorithm \cite{abk}. Although these
constructions
are at special points of the moduli space $(T^A_0,~U^A_0)$, the
generalization of them  for arbitrary  moduli is automatic by a
simple replacement of the fermionic "twisted" characters with the
characters of the "shifted" and "twisted" bosonic coordinates:
\be
Z_A\left[^{\gamma_A
{}~;~H_A}_{\delta_A~;~G_A}\right]|_{T^A_0,U^A_0}~\rightarrow
Z_A\left[^{\gamma_A ~;~H_A}_{\delta_A~;~G_A}\right]|_{T,U}.
\ee
Many models can be constructed in this way. We will display below the
partition function of a  model with one unbroken and three
spontaneously broken supersymmetries, $N=4\rightarrow N=1$ ( the
unbroken gauge group of this example is $E_8\times E_6 \times
U(1)^2$:
\ba
\Z^{4\rightarrow 1}(F_i)&=&{(\im)^{-1} \over \eta^2{\bar \eta}^2}
\times \frac{1}{4}\sum_{h_i,g_i}
Z_1\left[^{h_1 ~;~h_2}_{g_1~;~g_2}\right]~
Z_2\left[^{h_2 ~;~h_3}_{g_2~;~g_3}\right]~
Z_3\left[^{h_3 ~;~h_1}_{g_3~;~g_1}\right] \cr
&\times& {1\over 2}~\sum_{\alpha, \beta}~(-)^{\alpha+\beta
+\alpha \beta} ~\frac{\th[^{\alpha}_{\beta}]}{\eta}\frac{\th[^{\alpha
+
h_2}_{\beta + g_2}]}{\eta}
\frac{\th[^{\alpha + h_3}_{\beta + g_3}]}{\eta} \frac{\th[^{\alpha +
h_
1}_{\beta + g_1}]}{\eta} \cr
&\times& \frac{1}{2}~\sum_{\bar \alpha, \bar \beta}
\frac{{\bar \th}^{5}[^{\bar \alpha}_{\bar \beta}]}{{\bar \eta}^5}
{}~\frac{{\bar \th}[^{{\bar \alpha} + h_2}_{{\bar
\beta} + g_2}]}{{\bar \eta}}
{}~\frac{{\bar \th}[^{{\bar \alpha} + h_3}_{{\bar
\beta}~+~g_3}]}{{\bar \eta}}
 ~\frac{{\bar \th}[^{{\bar \alpha} + h_1}_{{\bar
\beta} + g_1}]}{{\bar \eta}} \times\cr
&&\delta(h_1+h_2+h_3)~\delta (g_1+g_2+g_3)\frac{1}{2}\sum_{\epsilon,
\zeta}\frac{{\bar
\th}^{8}[^{\epsilon}_{\zeta}]}{{\bar \eta}^8} .
\label{c5}\ea
The existence of one unbroken supersymmetry is ensured because of the
 relations $h_1+h_2+h_3=0$ and $g_1+~g_2+~g_3=0$; these relations
guarantee the
existence of an $N=2$ superconformal symmetry on the world-sheet and
thus the existence of $N=1$ space-time supersymmetry\cite{bd}.

 It is easy to see that the partition function $\Z^{4\rightarrow 1}$
can be decomposed in four sectors:

$\bullet$ {\bf The $N=4$ sector}, with  no rotations  or translations
in
all three complex planes  ($(h_A,~g_A)=(0,0)$)

$\bullet$ {\bf Three  $N=2$ sectors}, with a non-zero translation in
one of the complex planes and  opposite non-zero rotations  in the
remaining two complex planes.

The contribution to the partition function of the $N=4$ sector is one
quarter of the $N=4$ partition function with  lattice momenta in the
reduced $\Gamma (2,2)^3$ lattice.
The contribution of the other three $N=2$ sectors are equal sector by
sector
to the corresponding $N=4 \rightarrow N=2$ partition function divided
by a factror of 2. The  untwisted complex plane lattice momenta
correspond to the shifted  $\Gamma_{(2,2)}~[^{\gamma_A}_{\delta_A}]$
lattice.
The moduli-dependent corrections to the gauge couplings can be easily
determined by combining the results of the individual $N=2$ sectors.
\be
{16 \pi^2 \over g^2_{E_8}}~-{16 \pi^2 \over
g^2_{E_6}}~=~\Delta_{(8,6)}={1\over 2}\sum^3_{A=1}~\Delta^A_{(8,7)},
\label{c7}\ee
where the expressions of the $\Delta^A_{(8,7)}$ are given in
(\ref{b33}).

 As we  mentioned in the $N=4\rightarrow N=2$ spontaneous breaking,
one expects
a restoration of the $N=4$ supersymmetry in the limit in which the
massive gravitinos become massless; in order to prove the $N=4$
restoration in the $N=4\rightarrow N=1$ defined above  as a
$Z^2\times Z^2$ freely acting orbifold, we need to identify the three
massive gravitinos and express their masses in terms of the moduli
fields and the  three  $R$-symmetry charges $q_i~(i=1,2,3)$:
\be
m^2_{3/2}(q_i)~=~\frac{|q_2~-~q_3|^2}{4~{\rm Im}~T_1~{\rm Im}~U_1}~+~
\frac{|q_3~-~q_1|^2}{4~{\rm Im}~T_2~{\rm Im}~U_2}~+~
+~\frac{|q_1~-~q_2|^2}{4~{\rm Im}~T_3~{\rm Im}~U_3}~~
\label{c8}\ee
with $|q_0~+~q_1~+~q_2~+~q_3|=1~~{ \rm and}
{}~~|q_i|~=~|q_0|~=~{1 \over 2}$
where  $q_0$ is the left-helicity charge. Using the above
expression, one finds the desired result:
$$
(m^2_{3/2})_1~=~\frac{1}{4~{\rm Im}~T_2~{\rm
Im}~U_2}~+~\frac{1}{4~{\rm Im}~T_3~{\rm Im}~U_3}, ~~~~~
$$
$$
(m^2_{3/2})_2~=~\frac{1}{4~{\rm Im}~T_3~{\rm
Im}~U_3}~+~\frac{1}{4~{\rm Im}~T_1~{\rm Im}~U_1}, ~~~~~
$$
$$
(m^2_{3/2})_3~=~\frac{1}{4~{\rm Im}~T_1~{\rm
Im}~U_1}~+~\frac{1}{4~{\rm Im}~T_2~{\rm Im}~U_2}, ~~~~~
$$
\be
(m^2_{3/2})_0~=~0.
\label{c9}\ee
The three massive gravitinos become massless in the
decompactification limit
${\rm Im}~T_I~{\rm Im}~U_I $ $\to \infty$, $~I=1,2,3$, with ratios
${\rm Im}~T_I/{\rm Im}~U_I$ fixed. Thus the full restoration of the
$N=4$  effectively takes place in seven dimensions.
Partial restoration of an $N=2$ supersymmetry can happen in six
dimensions when ${\rm Im}T_I~{\rm Im}~U_I$ $\to\infty, ~~I=1,2$; In
this limit
$(m^2_{3/2})_0 ~=~ 0$ and  $(m^2_{3/2})_3 \rightarrow 0$.

\section{$N=2 \rightarrow N=1$  spontaneous SUSY breaking }
\setcounter{equation}{0}

Using similar techniques as before, it is possible to construct $N=2$
models  with one of the supersymmetries to be spontaneously broken,
$N=2\rightarrow N=1$. In this class of models the restoration of
$N=2$ takes place in six dimensions. No further restoration of
supersymmetry is possible. Examples
can be obtained as in $(T^2\otimes K_3)/ Z^2_{f}$ orbifold
compactification in which the $Z^2_{f}$ is freely acting. A
representative example of this class of models is the one in which
the $K_3$ compactification is chosen to be at the  orbifold point
$T^4/Z_o^2\sim K_3$  (we denote by  $Z^2_o$ the orbifold group and by
$Z^2_f$ that which corresponds to the freely acting orbifold). We
will give below  the exact partition function that corresponds to
this construction. From the  explicit expression we can directly
verify  the effective restoration of $N=2$ supersymmetry in the
large-volume limit of $K_3$. Using the $Z^2_o \otimes Z^2_f$ orbifold
notation, the partition function of the $( T^2 \otimes
T^4/Z_o^2)/Z^2_{f}$ model is:
\ba
\Z^{2\rightarrow 1}(F_i)&=&{(\im)^{-1} \over \eta^2
{\bar \eta}^2} \times   
\frac{1}{2}\sum_{h_f,g_f}\frac{1}{2}\sum_{h_o,g_o}
Z_1\left[^{0 ; h_f}_{0 ; g_f}\right]
Z_2\left[^{h_f  ;h_o}_{g_f ; g_o}\right]
Z_3\left[^{h_f ;-h_f-h_o}_{ g_f ; -g_f-g_o}\right] \cr
&\times&
{1\over 2} \sum_{\alpha, \beta} (-)^{\alpha+\beta
+\alpha
\beta} \frac{\th[^{\alpha}_{\beta}]}{\eta}\frac{\th[^{\alpha+
h_f}_{\beta + g_f}]}{\eta}
\frac{\th[^{\alpha + h_o}_{\beta + g_o}]}{\eta}
\frac{\th[^{\alpha -h_f-h_o}_{\beta -h_f-h_o}]}{\eta} \cr
&\times&
\frac{1}{2}~\sum_{\bar \alpha, \bar \beta}
\frac{{\bar \th}^{5}[^{\bar \alpha}_{\bar \beta}]}{{\bar \eta}^5}
\frac{{\bar \th}[^{{\bar \alpha} + h_f}_{{\bar
\beta} +g_f}]}{{\bar \eta}}
\frac{{\bar \th}[^{{\bar \alpha} + h_o}_{{\bar \beta} + g_o}]}{{\bar
\eta}}
\frac{{\bar \th}[^{{\bar \alpha} - h_f - h_o}_{{\bar
\beta}-g_f - g_o}]({\bar \tau})}{{\bar \eta}} 
\times
\frac{1}{2}~\sum_{\epsilon, \zeta}~\frac{{\bar
\th}^{8}[^{\epsilon}_{\zeta}]}{{\bar \eta}^8}.
\label{c10}\ea
In the above expression, the  parameters $(h_f,~g_f)$ and
$(h_o,~g_o)$    correspond to  $Z^2_f$ and  $Z^2_o$ respectively. The
unbroken gauge group of this model is the $E_8\otimes E_6 \otimes
U(1)^2$. Switching on continuous or discrete Wilson lines, we can
construct a large class of models with different gauge group but with
a universal behaviour
with respect to the $N=2$  restoration at the large moduli limit; the
 massive gravitino of the broken $N=2$ becomes massless when (${\rm
Im}~T_2~{\rm Im} U_2$ and ${\rm Im}~T_3~{\rm Im}~U_3~$ large).
\be
(m^2_{3/2})_1~=~\frac{1}{4~{\rm
Im}~T_2~{\rm Im}~U_2}~+~\frac{1}{4~{\rm Im}~T_3~{\rm Im}~U_3},
(m^2_{3/2})_0~=~0.
\label{c11}\ee
An easy way to view this ground-state is as an orbifold of the
original $N=4$ theory by the following non-trivial $Z_2\times Z_2$
elements: $(1,r,r), ~(r,rt,t), ~(r,t,rt)$, ($r$ stands for
``$\pi$-rotation" and $t$ for 1/2--lattice translation);
$(1,r,r)$ has four fixed planes while the others have none.
Because of  the $N=2$ restoration phenomenon, we expect that the only
non-vanishing
corrections to the gauge coupling constants are those that correspond
to the
$N=2$ sector with $(h_f,~g_f)=(0,~0)$ and $(h_o,~g_o)\ne (0,~0)$.
Indeed in this sector the $Z^2_o$ acts trivially  on the
$\Gamma_{(2,2)}(T_1,~U_1)$ lattice as in the usual orbifolds. On the
other hand, in the  remaining two $N=2$ sectors,

i) $(h_o, g_0)= (0, 0)$,  $(h_f, g_f)\ne(0, 0)$

ii) $(h_o, g_0)+(h_f,g_f)= (0, 0)$, $(h_f, g_f)\ne(0, 0)$.

In both sectors the corresponding  $Z^2$ acts without fixed points
because of the simultaneous non-trivial shift $(h_f,~g_f)$ on the
corresponding $\Gamma_{(2,2)}(T_A,~U_A), ~~A=2,3$, lattice.

The moduli-dependent corrections to the gauge couplings can be easily
determined by combining the results of the individual $N=2$ sectors.
\be
\Delta_{(8,6)}~=~{16 \pi^2 \over g^2_{E_8}}~-{16 \pi^2 \over
g^2_{E_6}}~=~ 
{1\over 2} \left(~\Delta^1_{(8,7)} ~+~ \Delta^2_{(8,7)} ~+~
{}~\Delta^3_{(8,7)}\right) ,
\label{c12}\ee
where the $\Delta^A_{(8,7)}$ are the threshold corrections of the
three $N=2$ sectors:
$$
\Delta^1_{(8,7)}=(b^1_8-b^1_7)~{\rm log}\left[ |\mu|^2{\rm Im}
T_1{\rm Im} U_1 |\eta(T_1)\eta(U_1)|^4 \right]
$$
$$
\rightarrow (b^1_8-b^1_7) \left[{\pi \over 3}(~{\rm
Im}T_1 + {\rm Im}U_1) + {\rm log} |\mu|^2 {\rm Im} T_1 {\rm Im}
U_1\right]
$$
\be ~\label{c13}\ee
which corresponds to the threshold corrections of the standard
orbifolds.

On the other hand $\Delta^A_{(8,7)}$ for $A=2,3$ will correspond to
the threshold corrections of freely acting orbifolds which have
different  behaviour in the large-moduli limit:
\ba
\Delta^A_{(8,7)}&=&(b^A_8-b^A_7)~{\rm log}~\left[ |\mu|^2~{\rm
Im}T_A~{\rm Im}U_A~\right]  
+~ (b^A_8-b^A_7){\rm log}~\left[ |\th_4(T_A)~\th_4(U_A)|^4 ~\right]~
\cr
&\rightarrow&(b_8-b_7)~{\rm log}~|\mu|^2~{\rm Im}T_A~{\rm Im}U_A~.
\label{c14}\ea
Modulo the artificial sub-leading logarithmic contribution (due to
the infrared divergences), the moduli contribution of the second and
third plane $T_A,~U_A,~ A=2,3$, is exponentially suppressed due to
the asymptotic behaviour of $\th_4(T_A),~\th_4(U_A)$ for large $T_A$
and $U_A$,
$ \th_4(T_A~)=1+{\rm \cal O}(e^{-i\pi~T_A })$.

There is a large class of such models obtained from $N=2$ $Z_2$
orbifold compactifications
by using $D_4$ type symmetries that act on the twist fields as well
as the lattice.

\section{Non-Perturbative BPS $N=4$ Mass Formula}
\setcounter{equation}{0}

The low-energy effective $N=4$ supergravity is  manifestly
invariant under the
$SO(6,22;R)$ group. The full string theory, however, is only
invariant under
the discrete subgroup $O(6,22,Z)$.
Furthermore the equations of motion and Bianchi identities are
invariant under  the $SL(2,R)$ transformation
\be
S \to {aS+b\over cS+d}~~~{\rm with}~~~ ad-bc=1,
\ee
 provided we perform an $SL(2,R)$ transformation on the ``electric"
and ``magnetic" fields and charges.
In particular, the transformation $S\to -1/S$ interchanges electric
and magnetic charges.
    It has been conjectured  \cite{S}--\cite{ss}
    that a discrete subgroup $SL(2,Z)$ of this continuous symmetry of
the equations of motion of the effective theory is a
(non-perturbative) symmetry of the full theory.In the heterotic
theory, only electrically charged states exist; these  charges are
the six  quantized momenta and windings as well as the 16 $U(1)$
charges of $E_8\times E_8$ or $SO(32)$, ($m_I,~n^I,~Q^k$). Obviously
the  spectrum of the heterotic theory is not invariant under
$SL(2,Z)$. For this to be true it is necessary to include in the
theory non-perturbative states that carry both electric
($m_I,~n^I,~Q^k$) and magnetic (${\tilde m}_I,~{\tilde n}^I,~{\tilde
Q}^k$) charges \cite{sen},\cite{cv}--\cite{CC}.
Thanks to the $N=4$ supersymmetric algebra and its central extension,
one can write down an exact mass formula for all stable perturbative
and non perturbative states, which preserves at least one of the four
supersymmetries (BPS-states).
\be
M_{BPS}^2~=~{(P_I+S~\Pi_I)~g^{IJ}~(P_J+\bar S~\Pi_J) \over 4{\rm Im}S} ~+~
{1 \over 2}
\sqrt{(P_I~g^{IJ}~P_J)~(\Pi_I~g^{IJ}~\Pi_J)~-~(P_I~g^{IJ}~\Pi_J)^2}
\ee
where the ``electric" and the ``magnetic" momenta $P_I$ and $\Pi_I$
are given in terms of
the ``electric" ($m_I,~n^I,~Q^k$) and ``magnetic" (${\tilde
m}_I,~{\tilde n}^I,~{\tilde Q}^k$) charges:
$$
P_I=
m_I + Y_I^k Q^k + {1\over2}Y_I^k Y^k_J n^J+ B_{IJ}n^J+G_{IJ}n^J
$$
$$
\Pi_I =
{\tilde m}_I + Y_I^k {\tilde Q}^k + {1\over2}Y_I^k Y^k_J {\tilde
n}^J + B_{IJ} {\tilde n}^J + G_{IJ} {\tilde n}^J
$$
\be ~ \ee

The square-root factor in the BPS mass formula  is proportional to
the difference of the two
central charges squared: depending on whether  this vanishes
or not, the representation
preserves  1/2  or  1/4 of the supersymmetries, (either short or
intermediate supermultiplets).
For the  perturbative BPS
states of the heterotic string
 (${\tilde m}_I,~{\tilde n}^I,~{\tilde Q}^k$) = 0, and thus belong to
short supermultiplets.
 Their  mass  reads

\be
M^2_{BPS, {\rm pert}}~= ~
{1\over
4 {\rm Im}S}~P_I~g^{IJ}~P_J.
\label{a15}\ee
The factor of ${\rm Im}S$ is there because masses are
measured in units of $M_{\rm Planck}$.
The BPS mass formula is manifestly invariant under
$SL(2,Z)_S$.
$$
S \rightarrow S+1;~~~~~P_I\rightarrow P_I~+~\Pi_I,~~~~\Pi_I
\rightarrow \Pi_I
{}~~~~~~~~~~~~~ ~~~~~~~~~
$$
\be
S \rightarrow -{1\over S};~~~~\Pi_I \rightarrow P_I,~~~~
P_I \rightarrow -\Pi_I.
\ee
Although the mass formula for non-perturbative BPS states is
understood, we do not know a priori the multiplicities of all these
states.
{}From the $N=4$ heterotic string we know the multiplicities when
$\Pi_I=0$. Using $SL(2,Z)$ we also know the multiplicities of all
states with
$P_I~g^{IJ}~\Pi_J=~0$. To go further and learn more about the states
with $P_I~g^{IJ}~\Pi_J~\ne ~0$ (namely intermediate multiplets) it is
necessary to go beyond the string picture and learn more about the
non-perturbative structure of the theory.
The heterotic string on $T^6$ is supposed to be equivalent, in the
strong coupling limit to the type II theory compactified on
$K_3\times T^2$.
Moreover, there is a hypothetical 11-d theory ($M$-theory) that
includes the
non-perturbative dynamics of type IIA theory \cite{wi}.
Thus compactification of $M$-theory on $K_3\times T^3$ contains all
the relevant non perturbative information about the heterotic $N=4$
theory.
This idea led to a  conjecture on the multiplicities of dyonic BPS
states in the 4-d $N=4$ theory \cite{dvv}.
This will be an important input, for our non-perturbative analysis of
the spontaneously broken $N=4$ theory.

\section{BPS states in models  with partial SUSY
breaking $N=4\to N=2$ }
\setcounter{equation}{0}

\def\l{\lambda}
Let us consider an interesting question concerning the BPS spectrum
of the theories where $N=4$ is spontaneously broken to $N=2$.
In the original heterotic $N=4$ theory, there are only short BPS
multiplets
in the perturbative spectrum.
Their multiplicities can be easily counted by using helicity
supertrace formulae
\cite{bk}.
In particular, the supertrace of helicity to the power 4 counts
the multiplicities of $N=4$ short (massless or massive) multiplets.
{}From the partition function of the heterotic $N=4$, we can
construct
the helicity-generating partition function:
\ba
\Z^{het}_{N=4}(v,\bar v)&=&{\rm Str}[q^{L_0}\;\bar q^{\bar L_0}e^{2\pi
iv\l_R-2\pi i\bar v\l_L}] 
={1\over
2}\sum_{\a\b}(-1)^{\a+\b+\a\b}~{\vartheta[^{\a}_{\b}](v)
\vartheta^3[^{\a}_{\b}]\over\eta^{12}\bar
\eta^{24}}~ \xi(v)~\bar\xi(\bar v)~ {\Gamma_{(6,22)}\over {\rm
Im}\tau} \cr
&=&{\vartheta^4_1(v/2)\over
\eta^{12}\bar \eta^{24}}\xi(v)\bar\xi(\bar v)
{\Gamma_{(6,22)}\over {\rm Im}\tau}
\label{cc18}\ea
The physical helicity in closed string theory $\l$ is the sum of the
left helicity $\l_L$ and the right helicity $\l_R$:
\be
\xi(v)~=~\prod_{n=1}^{\infty}{(1-q^n)^2\over (1-q^ne^{2\pi iv})
(1-q^ne^{-2\pi iv})}~=
{\sin\pi v\over \pi}{\vartheta_1'\over
\vartheta_1(v)}, \;\;\;\,\;\;\; (~\xi(v)=\xi(-v)~),
\label{cc3}\ee
which counts the contributions to the helicity due to the world-sheet
bosons.
If we define
\be
Q={1\over 2\pi i}{\partial\over \partial v}\;\;\;,\;\;\;\bar
Q=-{1\over 2\pi i}{\partial\over \partial \bar v},
\label{cc4}\ee
then
$$
B_4=\langle \lambda^4\rangle =(Q+\bar Q)^4 Z^{het}_{N=4}(v,\bar
v)|_{v=\bar v=0}={3\over 2}{\Gamma_{6,22}\over \bar\eta^{24}}.
$$
\be ~ \label{bb39}\ee
The numerator provides the mass formula while the denominator
$1/\bar\eta^{24}$ provides the multiplicities.
More precisely defined:
\be
{1\over \eta^{24}}=\sum_{N=-1}^{\infty}d(N)q^{N}={1\over q}+24 +324
q+{\cal O}(q^2).
\label{bb40}\ee
Then at the mass levels $M^2={1\over 4}P_L^2$, with
\be
q_e^2\equiv
2\vec m\cdot\vec n-\vec Q\cdot \vec Q,
\label{bb41}\ee
the multiplicity is $d(q_e^2/2)$.
The generalization of the string  ``electric"  multiplicity to
the non-perturbative  dyonic states needs to assume  a genus-2
generating function\cite{dvv}
$\Phi(\tau_{ij})$ ($\tau^{ij}=\tau^{ji}$):
\be
\Phi(\tau_{ij})~=~\sum_{N_{ij}}~d(N_{ij})~{\rm exp}
{}~[~{2i\pi~N_{ij}\tau^{ij}}~],
\label{phi}\ee
where the levels $N_{ij}$ of a dyonic state are characterized by  the
 electric and magnetic charges
$$
 \vec q_1\equiv \vec q_e =(\vec m,~\vec n,~ \vec Q)
,~~~~~~~~\vec q_2\equiv \vec q_m =(\vec {\tilde m},\vec {\tilde
n},\vec
{\tilde Q}) ~~~~~
$$
\be
2N_{ij}=\vec q_i \cdot \vec q_i~;
\label{bb42}\ee
the above equation  generalizes the ``electric" matching condition
for the dyons and magnetic monopoles. The non-perturbative
multiplicities are determined  in terms of the electric and magnetic
charges $d(\vec q_i \cdot \vec q_i /2))$.
Thus the knowledge of the generating function $\Phi(\tau_{ij})$
determines the full spectrum of the perturbative and non-perturbative
BPS states  in terms  of the moduli fields $(S,~T_{IJ},~Y^k_I)$ and
the charges.
In ref.~\cite{dvv}, it was conjectured that the generating function
$\Phi(\tau_{ij})$ is the genus-2 determinant of 24 bosons:
\be
\Phi(\tau_{ij})~=~\eta[\tau_{ij}]^{-24}~=~
\left[\prod_{even}~\th[\tau_{ij}]\right]^{-2}.
\ee

Using the  genus-2 interpretation of the non-perturbative
multiplicities,
we will present in  section-{\bf 9}  an algorithm  that  determines
the non-perturbative BPS spectrum of the $N=2$ theories in terms of
the shifted  $N=4$ spectrum. Here we will restrict ourselves to the
perturbative heterotic and type II cases. Some general facts, valid
in $N=2$ theories, are in order:

$\bullet$ The $N=2$ massless multiplets $M_0^{\lambda}$ have the
following helicity content:
\be
\pm\left(\lambda\pm{1\over 2}\right)+2(\pm \lambda);
\label{bbb43}\ee
$M_0^0$ is the hypermultiplet, $M_0^{1/2}$ is the vector multiplet,
while $M_0^{3/2}$ is the supergravity multiplet.

$\bullet$ The massive BPS multiplets have the following $SO(3)$ spin
content
\be
M^j\;\;:\;\;[j]\otimes ([1/2]+2[0])
\label{bbb441}\ee
and contain $2(2j+1)$ bosonic states and an equal number of fermionic
ones.

$\bullet$ Finally the generic long-massive multiplet has the
following SO(3) content
\be
L^j\;\;:\;\;[j]\otimes ([1]+4[1/2]+5[0])
\label{bbb451}\ee

$\bullet$ The $N=2$ BPS states correspond to the short multiplets
and  are picked up
by the supertrace of helicity squared, $B_2=\langle\lambda^2\rangle$.
We have
\be
B_2(M_0^{\lambda})=(-1)^{2\lambda}~,~B_2(L^j)=0~,~
B_{2}(M^j)=(-1)^{2j+1}(
2j+1)/2 .
\label{bbb461}\ee
For the  perturbative heterotic theory, a direct computation
determines  $B_2$ in terms of the characters of the shifted
$\Gamma_{2,18}[^h_g]$  and  of the four twisted (right-moving) bosons
$Z_{4,4}\left[^h_g\right]$. In what follows we will assume  for
simplicity the $Z_2$ freely acting orbifold  defined in section-{\bf
3}:
\ba
\tau_2~B_2&=& \tau_2~\langle \l^2\rangle
 =  \Gamma_{2,18}[^0_1]{\bar\th^2_3\bar\th_4^2\over
\bar\eta^{24}}-\Gamma_{2,18}[^1_0]{\bar\th^2_2\bar\th_3^2\over
\bar\eta^{24}}-\Gamma_{2,18}[^1_1]{\bar\th^2_2\bar\th_4^2\over
\bar\eta^{24}} \cr
&=& {\Gamma_{2,18}[^0_0]+\Gamma_{2,18}[^0_1]\over 2}\bar F_1
-{\Gamma_{2,18}[^0_0]-\Gamma_{2,18}[^0_1]\over 2}\bar F_1 \cr
&-&{\Gamma_{2,18}[^1_0]+\Gamma_{2,18}[^1_1]\over 2}\bar
F_+
-{\Gamma_{2,18}[^1_0]-\Gamma_{2,18}[^1_1]\over 2}\bar F_- ~.
\label{bb44}\ea
with
\be
\bar F_1={\bar\th^2_3\bar\th_4^2\over \bar\eta^{24}}\;\;\;,\;\;\;
\bar F_{\pm}={\bar\th^2_2(\bar\th_3^2\pm\bar\th_4^2)\over
\bar\eta^{24}}.
\label{bb45}\ee
For all $N=2$ heterotic theories $B_2$ has universal modular
properties under $SL(2,Z)_{\tau}$:
\be
\tau\to\tau+1~~~~B_2\to B_2~~~~
\tau\to -{1\over
\tau}~~ ~~
B_2\to \tau^2 ~B_2.
\label{bb46}\ee
All functions $\bar F_i$ have positive coefficients and have the
generic expansions
\be
F_{1}={1\over q}+\sum_{n=0}^{\infty}d_{1}(n)q^n 
=~ {1\over q}+16+156
q+{\cal O}(q^2)
\label{bb47}\ee
\be
F_{+}~=~ {8\over q^{3/4}}+q^{1/4}\sum_{n=0}^{\infty}d_{+}(n)q^n~=
{8\over
q^{3/4}}+8q^{1/4}(30+481 q+{\cal O}(q^2))
\label{bb48}\ee
\be
F_{-}~=~{32\over q^{1/4}}+q^{3/4}\sum_{n=0}^{\infty}d_{-}(n)q^n~=
{32\over q^{1/4}}+32q^{3/4}(26+375q+{\cal O}(q^2)).
\label{bb49}\ee
Also the lattice sums $(\Gamma_{2,18}[^h_0]\pm\Gamma_{2,18}[^h_1])/2$
have
positive multiplicities.

The contribution of the generic massless multiplets is given by the
constant coefficient of $F_1$ and agrees with the expectation,
$16=20-4$, since we have the supergravity multiplet and 19 vector
multiplets contributing $20$, and $4$ hypermultiplets contributing
$-4$.
Turning off all the Wilson lines and restoring the $E_7\times E_8$
group, the above result  becomes
\be
\langle ~ \l^2 ~\rangle
{}~ =~
\Gamma_{2,2}[^0_1]{\bar\th^4_3\bar\th_4^4(\bar\th_3^4+\bar\th_4^4)
\bar E_4\over 2\bar\eta^{24}}~+
-\Gamma_{2,2}[^1_0]
{\bar\th^4_2\bar\th_3^4(\bar\th_2^4+\bar\th_3^4)\bar E_4\over
2\bar\eta^{24}}-\Gamma_{2,2}[^1_1]{\bar\th^4_2\bar\th_4^4
(\bar\th_2^4-\bar\th_4^4)\bar E_4\over 2\bar\eta^{24}}
\label{bb45a}\ee.
Let us analyse the $N=4\to N=2$ BPS mass formulae;
we  denote by ${\vec\e}=(\vec\e_L;\vec\e_R,\vec\zeta)$ the shift
vector
of the $\Gamma_{(2,18)}$; ${\vec\e}$ must satisfy the
modular-invariant constraint ${{\vec\e}\cdot{\vec\e}}\equiv
2{\vec\e}_L\cdot\vec\varepsilon_R-{\vec\zeta\cdot\vec\zeta
}=2(-1+{\rm mod}~ 4)$.
 The mass formula  for the BPS states is:

$\bullet$  In the ``untwisted" sector $h=0, ~ {\vec\e}=0$\\
\vskip .1cm
\be
M^2(h=0)~ =
{|-m_1U+m_2+Tn_1+(TU-{1\over 2}
\vec W^2)n_2+\vec W\cdot \vec Q|^2\over 4 \;S_2
\left(T_2U_2-{1\over 2}{\rm
Im}\vec W^2\right)},
\label{bb50}\ee
where $\vec W$ is the 16-dimensional complex vector of Wilson lines.
When the integer
\be
\rho=\vec m\cdot \vec \e_R+\vec n\cdot\e_L-\vec Q\cdot\vec\zeta
\label{bb51}\ee
is even, these states are vector-like multiplets with multiplicity
function
$d_1(s)$ of (\ref{bb47}) where $s$ is :
\be
s=\vec m\cdot\vec n-{1\over 2}\vec Q\cdot\vec Q.
\label{bb52}\ee
When $\rho$ is odd, these states are hypermultiplets-like  with
multiplicities $d_1(s)$.

$\bullet$ In the ``twisted" sector $h=1$, ${\vec\e}\ne0$\\
\vskip .1cm
\be
M^2(h=1)~=
{|-m'_1U+ m'_2 + Tn'_1+(TU-{1\over 2}
\vec W^2)n'_2+\vec W\cdot \vec Q'|^2\over 4 \;S_2
\left(T_2U_2-{1\over 2}{\rm
Im}\vec W^2\right)}
\label{bb53}\ee
with
\be
\vec m' \equiv \vec m+{\vec \e_L\over 2},~~~\vec n'\equiv \vec
n+{\vec \e_R\over 2},~~~\vec Q'\equiv \vec Q+{\vec \zeta\over 2}.
\ee
When $\rho$ is even ($\rho'$ odd) the states  are hypermultiplet-like
with multiplicities
$d_+(s')$, with
\be
s'=\vec m' \cdot\vec n'-{1\over 2}\vec Q'\cdot\vec Q'
\label{bb54}\ee
and
\be
\rho'=\vec m'\cdot \vec \e_R+\vec n'\cdot\e_L-\vec Q'\cdot\vec\zeta,
\label{bb54b}\ee
When $\rho$ is odd ($\rho'$ even) the states  are hypermultiplet-
like with multiplicities
$d_-(s')$.

Let us discuss here the gauge-symmetry enhancements in the presence
of shift vectors.
For simplicity we will ignore the charged sector coupled to the
Wilson lines and focus on the $\gamma_{(2,2)}$ part.
Let us fist consider the untwisted sector $(h=0)$.
According to the above analysis, the masses are given by the
unshifted mass formula
(\ref{bb50}) and they are vector multiplets when  $\rho$ is even and
hypermultiplets when $\rho$ is odd.
Now the points where the standard $\Gamma_{(2,2)}$ mass vanishes are
well known.
At $T=U$, there are two configurations with zero mass,
given by $m_1=n_1=\pm 1$, all the rest being zero.
For both states, $|\rho|=|\e_L^1+\e_R^1|$. Depending on it being even
or odd, these states are either vector multiplets that enhance the
gauge group
$U(1)^2\to SU(2)\times U(1)$ or hypermultiplets charged under one of
the
$U(1)$'s.

Let us now look for states becoming massless in the twisted ($h=1$)
sector at $T=U$.
Again we obtain $m_1+\e_L^1/2=n_1+\e_R^1/2$. Since $\e^1_L,\e^1_R$
are either 0 or 1, the previous condition can be satisfied  only if
both $\e_L^1=\e_{R}^1=\psi$, with $\psi=0,1$.
Then $\rho=2m_1\psi$ and is always even.
The matching condition here for such a state becomes $s=3/4$ when
$\rho$ is even (see (\ref{bb48})).
Thus the condition on $m_1$ becomes
\be
m_{1}^2+\psi m_1={3\over 4}-{\e^2\over 8}.
\label{bb55}\ee
{}From modular invariance we have $\e^2/2=-1$ mod $4=4k-1$ $k\in Z$;
then eq. (\ref{bb55}) becomes
\be
m_1^2+\psi m_1+k-1=0
\label{bb56}\ee
and has either  two solutions or none in the field of integers,
depending on $\psi$ and $k$.
All such potentially massless states are hypermultiplets, come with
multiplicity 8 and have equal and opposite charge under one of the
2-torus
$U(1)$'s.

\section{Heterotic-Type II dual pairs with   partially
broken SUSY $N=4 \to N=2$}
\setcounter{equation}{0}

The heterotic string compactified on $T^4$, with $N=2$ in $(6-d)$
space-time supersymmetry, has been conjectured to be dual to type II
theory compactified on $K_3$ \cite{duff,HT}.
This duality changes the sign of the dilaton, dualizes the field
strength of the antisymmetric tensor and leaves the (4,20) gauge
fields $A^{I}_{\mu}$, the $SO(4,20)$ moduli and the Einstein metric
invariant.
Obviously this duality descends in four dimensions by compactifying
both theories on an extra $T^2$.
In four dimensions there are four extra gauge fields, two coming from
the metric $A^i_{\mu}$ whose charges are the momenta of the $T^2$ and
two coming from
the antisymmetric tensor $B_{i,\mu}$, whose charges are the winding
numbers of the $T^2$.
Also, we have three extra scalars from the components of the metric
on $T^2$, $G_{ij}$ and one from the antisymmetric tensor $B_{ij}$.
There are also 2$\times$ 24 extra scalars, $Y^i_I$ coming from the
6-d gauge bosons plus one more $A$, which is the  four-dimensional
dual of the antisymmetric tensor.
If we denote heterotic variables by unprimed names and type II ones
by primed
names, then the heterotic-type II duality in four dimensions implies
that

\be
e^{-\phi}=\sqrt{{\rm det}G'_{ij}}\;\;\;,\;\;\;
e^{-\phi'}=\sqrt{{\rm det}G_{ij}}
\label{d1}\ee
\be
{G_{ij}\over \sqrt{{\rm det}G_{ij}}}={G'_{ij}\over \sqrt{{\rm
det}G'_{ij}}}\;\;\;,\;\;\;A'^{i}_{\mu}=A^{i}_{\mu}
\label{d2}\ee
\be
e^{-\phi}g_{\mu\nu}=e^{-\phi'}g'_{\mu\nu}\;\;\to\;\;g^{E}_{\mu\nu}=
g'^{E}_{\mu\nu}
\label{d3}\ee
\be
M'_{4,20}=M_{4,20}\;\;\;,\;\;\;A^{I}_{\mu}=A'^{I}_{\mu}\;\;\;,\;\;\;
Y^{i}_{I}=Y'^{i}_{I}
\label{d4}\ee

\be
A={1\over 2}\e^{ij}B'_{ij}\;\;\;,\;\;\;
A'={1\over 2}\e^{ij}B_{ij}.
\label{d6}\ee
Moreover, it effects an electric--magnetic duality transformation on
the $B^i_{\mu}$ gauge fields
\be
{1\over 2}{{\e_{\mu\nu}}^{\rho\sigma}\over
\sqrt{-{\rm det}g}}\e^{ij}F^{B'}_{j,\rho\sigma}
={\delta S^{het}\over \delta F^{B,\mu\nu}_{i}}.
\label{d7}\ee
On the electric and magnetic charges it acts on the $T^2$
charges and leaves the rest invariant.

For the configurations of moduli we are interested in, namely
the factorization $(6,22)\to (2,18)\times (4,4)$, we proceed as
follows.
In the case of the heterotic string the complex moduli $T,U,\vec W$
are defined in terms of $G_{ij},~B_{ij}$ and $Y^k_i$, $i,j=(1,2)$.
However, for the type II string the situation is different.
A careful analysis of the tree-level action shows that there is an
analogue
of the Green--Schwarz term $B\wedge F\wedge F$ at tree level; this
appears
at one loop at the heterotic side for 4-d descendants of  both
$B\wedge F^4 $ and $B\wedge R^4$; the $B\wedge R\wedge R$ term
appears  at one loop in the type II side \cite{vw2}.
This term changes at tree level the definition of the type II $S'$
field.
There is an analogous phenomenon, which changes also at tree level
the definition of the $T'$ field.
The correct formulae read:
$$
S'=A'-{1\over 2}Y^I_1Y^I_2+{U_1\over
2}Y^I_2Y^I_2+i\left((e^{-\phi'}+{U_2\over 2}Y^I_2Y^I_2\right)
$$
\be ~ \label{e5}\ee
\be
T'=\sqrt{{\rm det}G'_{ij}}+iB'
\ee
where as usual
\be
{1\over \sqrt{{\rm det}~G'_{ij}}}G'_{ij}={1\over
U_2}\left(\matrix{1&U_1\cr
U_1&|U|^2\cr}\right).
\label{e6}\ee
Thus (\ref{d1})--(\ref{d6}) translate to
\be
U=U'\;\;\;,\;\;\;\vec W=\vec W'
\;\;\;,\;\;\;\
S=T'\;\;\;,\;\;\;T=S'.
\label{d14}\ee

Let us indicate how the $N=4$ heterotic-type II duality works at the
level the restricted $SO(2,18)$ BPS formula:
\be
M^2_{BPS}={|P~+~S~\Pi|^2 \over {\rm Im}S ( {\rm Im}T {\rm Im}U -
{1\over 2} {\rm Im}{\vec W}\cdot {\rm Im}{\vec W} )}
\label{n2bps}\ee
where $P$ and $\Pi$ are given in terms of the ``electric" and
``magnetic" charges and in terms of the complex moduli $T,U,\vec W$:
$$
P=-m_1+ n_1T + m_2 U + n_2(TU-{1\over 2}{\vec W}\cdot{\vec W}) +
Q\cdot{\vec W}
$$
and
$$
\Pi=-{\tilde m}_1+ {\tilde n}_1T+ {\tilde m}_2 U+ {\tilde n}_2
(TU-{1\over 2}{\vec W}\cdot{\vec W} ) + {\tilde Q}\cdot {\vec W}
$$
\be ~\label{n2bps1}\ee
We start first from the heterotic string not necessarily weakly
coupled.
We would like, however, to end up and compare with the weakly coupled
type II string.
Thus we must take the limit $T_2$ large in the mass formula and keep
light states:
\be
M^2_{het}~=
{|-m_1+m_2 U+\vec W\cdot Q+S(\tilde
m_{2} -\tilde m_{1}U+\vec W\cdot\tilde {\vec Q})|^2\over
4~S_2(T_2U_2-(\vec
W_2)^2/2)}
\label{e1}\ee
However the terms containing the charged states are really absent;
$\vec m\cdot \vec n=0$ and $\tilde {\vec m}\cdot \tilde {\vec n}=0 $
and thus, there are no physical states with non-trivial
$\vec Q,\tilde{\vec Q}$.
Taking this into account, then using type II variables from
(\ref{d14}), we can
write (\ref{e1}) as

$$
M^2_{pert-II}={|-m_1+m_2 U'+T'(\tilde m_{2})-\tilde m_{1}U'|^2\over
4~\left(S'_2-{\vec {W'}_2^2\over 2U'_2}\right)T'_2U'_2}
$$
\be ~\label{e2}\ee
which gives the correct tree-level type II mass formula in the large
$T'_2$ limit, taking into account (\ref{e5}) and the duality map.

 Owing to the adiabatic argument of ref.\cite{vw}, we can obtain
new dual heterotic--type II pairs by orbifolding both the $N=4$
heterotic and $N=4$ type II strings, by the same freely acting
symmetry.
Thus we would like to identify the duals of the heterotic models
constructed
in the previous sections with spontaneously broken supersymmetry.

For concreteness we will go to the $Z_2$ sub-manifold of $K_3$, where
the conformal field theory is explicit, and we will map directly the
heterotic to
the type II string.
The type II partition function on $K_3\times T^2$ at the orbifold
point is
\ba
Z^{II}_{N=4}&=& {1\over {\rm Im}\tau |\eta|^4} ~{1 \over
2}\sum_{h,g=0}^{1} {\Gamma_{(2,2)}\over
|\eta|^{4}}~Z_{(4,4)}^{\rm twist}[^h_g]
\times {1\over2} \sum_{\a,\b=0}^{1}(-1)^{\a+\b+\a\b}
{}~{\vartheta^2[^{\a}_{\b}]
\vartheta[^{\a+h}_{\b+g}]\vartheta[^{\a-h}_{\b-g}]\over
\eta^4}\cr
&\times& {1 \over 2}\sum_{\bar\a,\bar\b=0}^{1}
(-1)^{\bar\a+\bar\b+\bar\a\bar \b}
{}~{\bar
\vartheta^2[^{\bar\a}_{\bar\b}]
\bar\vartheta[^{\bar\a+h}_{\bar\b+g}]
\bar\vartheta[^{\bar\a-h}_{\bar\b-g}]\over \bar\eta^4}
\label{d9}\ea
Let us examine  the massless bosonic spectrum of the $N=4$ type II,
and try to match it to that of the $N=4$ heterotic string
\cite{{fk},{HT}}.

$\bullet$ In the NS--NS $(\a=\bar\a=0)$  untwisted sector $(h=0)$,
there are  32 degrees of freedom,
corresponding to the graviton, 2 scalars (axion-dilaton), 4 vectors,
and another 20 scalars
(the $\Gamma_{2,2}$ and $Z^{\rm twist}_{(4,4)}$ moduli).
Two of the gauge bosons are graviphotons while the other two belong
to $U(1)$ vector multiplets. Thus these four gauge bosons have
lattice signature (2,2).
Similarly the (2,2) moduli belong to these two vector multiplets
while the (4,4) moduli are in multiplets with vectors coming from the
R--R untwisted sector.

$\bullet$ In the NS--NS $(\a=\bar\a=0)$ twisted sector $(h=1)$, there
are 16 $Z_2$ invariant states in
the $T^4/Z_2$ part: $H^{I}$.
There are in total $4\times 16$ massless states;
all of them are scalars in multiplets with vectors coming from the
R-R twisted sector.

$\bullet$ In the R--R $(\a=\bar\a=1)$ untwisted sector there are 32
physical degrees of freedom.
These correspond to 8 vectors and 16 scalars.
The vectors have lattice signature (4,4) and four of them are
graviphotons
while the other four are in vector multiplets.
The sixteen scalars complete the six vector multiplets.

$\bullet$ In the R-R, twisted sector, there are $4\times 16$ massless
states corresponding to 16 vectors and 32 scalars.

Here the gauge group is composed of $U(1)$'s, which implies that we
are sitting
at a generic point in the space of Wilson lines.
The perturbative spectrum is charged under two of
the graviphotons and two of the other gauge bosons with charges given
by $P_L,P_R$ of the $T^2$.

Consider now the freely acting orbifold on the heterotic side acting
as
a $\pi$ rotation on the (4,4) part of the lattice
and as a translation $\vec\e$ on the $\Gamma_{(2,18)}$ lattice. Again
for simplicity we focus on the $Z_2$ case.  On the type II side the
$Z_2$ rotation on the (4,4) part changes the sign
of the massless states coming from the untwisted R--R sector as well
as the scalars coming from the twisted NS--NS sector.
The effect of the (2,18) translation
$\vec\e=(\vec\e_L;\vec\e_R,\vec\zeta)$ is to give phases to massive
charged states, but has no effect on the massless spectrum.
Thus at the massless level the NS--NS twisted and R--R untwisted
sectors have to be projected out.
The projection in the type II  case, which has the same effect as the
(4,4) rotation in the heterotic side, is a combination of the right
fermion number operator $(-1)^{F_{R}}$, which changes the sign of
the right-moving Ramond sector, and the symmetry transformation
$e=(-1)^h$, which acts on the twisted states of the orbifold
with a minus sign and is inert on anything else.

The $\vec\zeta$ translation vector does not act in the perturbative
type II string
since the perturbative spectrum does not contains states charged
under the 16
gauge bosons coming from the R-R twisted sector.
However it will act on non-perturbative $D$-brane states carrying
R--R charges.
Finally the phase coming from the translation of the (2,2) piece is
\be
(-1)^{\vec m\cdot\e_{R}+\vec n\cdot\e_{L}}
\label{d11}\ee
in the heterotic side.
Under the type II--heterotic map, this becomes, in the type II side:
\be
(-1)^{\vec m\cdot\vec\e_{R}+\tilde{\vec m}\hat\times\vec \e_{L}}
\label{d12}\ee
where $\vec a\hat\times \vec b=a_1b_2-a_2b_1$.
Thus the $\e_L$ translation acts on the type II side on the
magnetically charged states of the momentum-gauge fields of the
two-torus; it is  thus, not visible in type II perturbation theory.

The type II duals have 20 vector-multiplets and 4 hypermultiplets;
thus they are ``mirrors" of the  type II models  discussed in
ref.~\cite{vw}
with 4 vector multiplets and 20 hypermultiplets.
Therefore, the  perturbative partition function of the  type II
models dual to the heterotic ones is
\ba
Z_{II}^{4\to 2}&=&{1\over {\rm Im}\tau|\eta|^4}
{}~{1\over 2} \sum_{h,g,\bar h\bar
g=0}^{1} {\Gamma^{\vec \e_R}_{2,2}[^{\bar h}_{\bar g}]\over
|\eta|^{4}}Z_{(4,4)}^{\rm twist}[^h_g] \cr
&\times &{1\over 2}\sum_{\a,\b=0}^{1}~(-1)^{\a+\b+\a\b}~
{\vartheta^2[^{\a}_{\b}]
\vartheta[^{\a+h}_{\b+g}]\vartheta[^{\a-h}_{\b-g}]\over
\eta^4}
\times {1\over 2} \sum_{\bar\a,\bar\b=0}^{1}
(-1)^{\bar\a+\bar\b+\bar\a\bar \b}~{\bar
\vartheta^2[^{\bar\a}_{\bar\b}]
\bar\vartheta[^{\bar\a+h}_{\bar\b+g}]
\bar\vartheta[^{\bar\a-h}_{\bar\b-g}]\over \bar\eta^4}\cr
&\times& (-1)^{ (\bar\a+h)\bar g+(\bar
b+g)\bar h+\bar g\bar h}.
\label{d10}\ea
Here the reader might have noticed a potential puzzle.
Consider a heterotic model defined by a translation  vector with
$\vec\e_{R}=\vec 0$.
In this model, in the limit ${\rm Im}~T\to 0$ $N=2$ supersymmetry
is restored to $N=4$. Alternatively speaking, $m_{3/2}\sim {\rm Im}~
T$.
Thus in weakly-coupled heterotic string we take $S\to \infty$
and also $T\to 0$.
According to our duality map described above, there is no
perturbative
shift of the $T^2$ in the type II side.
Thus, at the perturbative level, the type II $N=2$ theory does not
look like
a spontaneously broken $N=4$.
However a look at (\ref{d12}) is sufficient to convince us that
there are two  gravitinos, with $m_{3/2}\sim {\rm Im}~S'$, which are
light in the strong coupling region of the type II theory and
certainly
not visible in the weak coupling type II perturbation theory.

A similar phenomenon can happen in reverse.
Consider a freely acting orbifold of the type II $(N=4)$ side, as in
(\ref{d10}),
where the (2,2) lattice translation acts on the windings of the
two-torus
with the phase $(-1)^{\vec\e_L\cdot \vec n}$.
This is modular-invariant on the type II side.
On the heterotic side the shift of the two-torus becomes
non-perturbative
via the heterotic-type II map, $(-1)^{\tilde{\vec m}\hat \times
\vec\e_{L}}$.
Thus, in heterotic perturbation theory, we only see the $Z_2$
rotation of
the (4,4) torus. As it stands the heterotic $N=2$ model is not
modular-invariant. An extra shift in the gauge lattice is needed (not
visible on
the type II side).
Thus the perturbative heterotic ground state has a $K_3\times T^2$
structure
(at the $Z_2$ orbifold point) and the supersymmetry $N=4\to N=2$
looks explicitly
broken in perturbation theory.
Turning on all Wilson lines we find that the generic massless
spectrum has 19 vector multiplets (including the dilaton) and 4
hypermultiplets.
Moreover the $SL(2,Z)_S$ is broken to $\Gamma^-(2)_S$ as can  easily
be seen by following the fate of $T$-duality of the type II dual.

Another comment concerns the fate of the $SL(2,Z)_S$
electric--magnetic duality symmetry of the original N=4 theory, in
the spontaneously broken phase.
It is known that in the $N=4$ case $SL(2,Z)_S$ is a corollary of
heterotic--type II duality, since the $T$-duality of type II
translates into the $S$-duality of the heterotic theory.
Let us investigate what remains of the perturbative $T$ duality in
the
broken type-II theory.
We have argued above that the two-torus on the type II side gets a
(perturbative) shift
$(\vec 0;\vec\e_R)$ that amounts to the phase $(-1)^{\vec
m\cdot\vec\e_{R}}$.
The $SL(2,Z)_T$ acts on the two-torus charges as the set of matrices
$$
SL(2,Z)_T ~:\;\;\left(\matrix{\vec m\cr\vec n\cr}\right)\to
\left(\matrix{a~{\bf 1}&b~i\sigma^2\cr -c~i\sigma^2&d~{\bf
1}\cr}\right)
\left(\matrix{\vec m\cr\vec
n\cr}\right);
$$
\be
ad-bc=1\;\;,\;\;a,b,c,d\in Z
\label{d13}\ee
There are two subgroups of $SL(2,Z)$ that are relevant here;
one is $\Gamma^+(2)$ defined by $b$ even in (\ref{d13}); the other
one is $\Gamma^-(2)$ defined by $c$ even in (\ref{d13}).
Thus when $\vec\e_R\not=\vec 0$, $SL(2,Z)_{T}$ is broken to
$\Gamma^+(2)_T$.
Thus, the $S$-duality group is reduced to $\Gamma^+(2)_S$.

In the above discussion , it is obvious that there are
non-perturbative
ambiguities in the translation-related projections.
The most general projection conceivable is determined by the
``electric"
translation vector $\vec\e$, but simultaneously by a ``magnetic"
translation vector $\tilde \e$ whose effects are not visible in the
perturbative spectrum.
Parts of these translations are never perturbatively visible either
in the heterotic nor in the type II side.
We will comment more on this issue in the next section.

One more remark is in order about the type II duals described above.
Inspection shows that all of the $N=2$ spacetime supersymmetry comes
from the left side. Consequently, in these models the $S$ field is in
a
vector multiplet \cite{vw}.
Thus, as in the heterotic side, the vector-moduli space gets
corrections while the hypermultiplet moduli space does not.
At generic Wilson lines this class of models has a massless spectrum,
which consists, apart from the supergravity and the dilaton vector
multiplet,
of 18  vector multiplets and 4 neutral hypermultiplets (the moduli of
the four-torus).
The non-perturbatively exact hypermultiplet quaternionic manifold is
$SO(4,4)/SO(4)\times SO(4)$.
The exactness of the hypermultiplet moduli space restricts the
orbifolding possibilities on the type II side to the ones described
in (\ref{d10}).

\section{Non-perturbative BPS spectrum in partially broken SUSY $N=4
\to N=2$ }
\setcounter{equation}{0}

Our conjecture for the non-perturbative multiplicities consists in
generalizing  the perturbative multiplicity functions
(\ref{bb47})--(\ref{bb49}) $F_i$ in genus-2.
First we rewrite $F_i$ in a more convenient form:
\be
F_1 =   {1 \over {\bar \eta}^{24}}~ \chi \left[^0_1\right],~~
F_{\pm}={1 \over {\bar \eta}^{24}}~
\left({\bar \chi} \left[^1_0\right]\pm{\bar \chi}
\left[^1_1\right]\right)
\ee
where ${\bar \chi}\left[^h_g\right]$ are given in terms of the
characters of four twisted 2d right-moving bosons:
\be
{\bar \chi}\left[^h_g\right]~
=~{4(-)^h~{\bar \eta}^6 \over {\bar
\th}[^{1+h}_{1+g}]~{\bar\th}[^{1-h}_{1-g}]},
\ee
where in the above equation $(h,g)\ne 0$. We can extend the validity
of ${\bar \chi}\left[^h_g\right]$  for all  $(h,g)$ sectors using
identities between right-moving,  bosonic and fermionic, ``twisted"
characters:

${\bar \chi}\left[^h_g\right]~=$
\be
{1\over 8~{\bar
\eta}^6}~\sum_{a,b}(-)^h~{{\bar\th}^4[^{a+h}_{b+g}]~
{\bar\th}^4[^{a-h}_{b-g}]~
{\bar\th}[^{1+h}_{1+g}]~
{\bar\th}[^{1-h}_{1-g}]}.
\label{char}\ee
In this expression, the absence of the $(h,g)=(0,0)$ sector is due to
the vanishing of the odd-spin structures (${\bar\th}[^1_1]$ terms).
In genus-2
$h$ and $g$ become ${\vec h}=(h,~{\tilde h})$ and ${\vec
g}=(g,~{\tilde g})$ in correspondence with the ``electric" and
``magnetic" charge shifts. The generalization in genus-2 of the
twisted characters consists in promoting
the various $\th$-functions with characteristics in genus-2
\be
{\bar\th}[^{a~+~h}_{b~+~g}]({\bar \tau})~~~\to~~~ {
\bar\th}\left[^{{\vec
a}~+~{\vec h}}_{{\vec b}~+~{\vec g}}\right]({\bar \tau}^{ij}).
\ee
Then, the proposed non-perturbative multiplicities will be generated
by the genus-2  functions:
\be
F \left[^{\vec h}_{\vec g}\right]~=~{ \Phi({\bar \tau}^{ij})}~{\bar
\chi}\left[^{\vec h}_{\vec g}\right]({\bar\tau}^{ij}),
\ee
where $\Phi({\bar \tau}^{ij})$ is the $N=4$ multiplicity function and
 ${\bar \chi}\left[^{\vec h}_{\vec g}\right]({\bar \tau}^{ij})$ are
the genus--2 analogues of the genus-1 ``twisted" characters
${\bar \chi}[^h_g]({\bar \tau})$ defined above.

Using the genus-2 multiplicity functions, we can construct  weighted
free-energy super-traces, which extend at the  non-perturbative level
the same perturbative quantities, e.g. the moduli dependence of the
gauge and gravitational couplings. We define by ${\cal L}^{\cal D}$
the following quantity:
\ba
{\cal L}^{\cal D}&=&\int_{\cal C} [dt\prod
dX^{ij}]~\sum_{h_i,g_i}~\sum_{q_i}~{\cal D}(\tau^{ij})~
F \left[^{\vec h}_{\vec g} \right] ({\bar \tau}^{ij})~
\times ~{\rm exp}~\left[-2i\pi~{\rm Re}{\tau} ^{ij}~({\vec q}_i+{\vec
\e}_i)\cdot({\vec
q}_j+{\vec \e}_j) ~\right]\cr
&\times&{\rm exp}~\left[-\pi~t~M^2_{BPS}(S;~{\vec q}_i,{\vec \e}_i)~
 \right],
\ea
where $M^2_{BPS}(S;~{\vec q}_i,{\vec \e}_i)$ stands for the
non-perturbative
mass formula (\ref{n2bps},\ref{n2bps1} with shifted charges;
$M^2_{BPS}$  depends on the  shifted ``electric" and ``magnetic"
charges, the moduli $T, U,$ and ${\vec W}$ as well as the
dilaton--axion  moduli field $S$.
 The period  matrix $\tau^{ij}$ of genus-2 in eq.(\ref{tauij}),  is
constructed in terms of  the  parameters $t$, $X^{ij}$ and  $S$ in
the following way:
\be
t=\sqrt{{\rm det}(\tau^{ij})},~
X^{ij}={\rm Re}~\tau^{ij},~{\rm and}~
{ \tau^{ij}\over \sqrt{det~\tau^{ij}} }={1\over {\rm
Im}S}\left(\matrix{1&{\rm Re}S\cr {\rm Re}S&|S|^2\cr}\right).
\label{tauij}\ee
The integration on $X^{ij}$ in the domain $[-1/2,~+1/2]$ would give
rise to the non-perturbative matching conditions (\ref{bb42}).
The  relevant multiplicities are generated by the functions  $F
\left[^{\vec h}_{\vec g} \right]$.
This is a suggestive argument, and stands in a similar footing with
the analogous $\tau_1$ integration in the perturbative string.
However we suggest that, like in the string case,  the correct
integration domain is the genus-two  fundamental region.
Thus we expect that the integration over $t$ (in the fundamental
domain of genus-2 with $S$ fixed) gives rise to the non-perturbative
quantity ${\cal L}^{\cal D}[S;~T,~U,~{\vec W}]$ in terms of all
moduli, $S$ included.

The kernel ${\cal D}$ is the  genus-2 analogue of a product of charge
operators. In the perturbative string, this is given by a product of
right-moving lattice vectors and contains also a ``back-reaction"
term \cite{kk}.
There is an analogue of ``right-moving" charges in the
non-perturbative
case
when we also include the magnetic charges.
The charge sum for the overall trace can be written in the
perturbative case as a $\bar\tau$ derivative, which generalizes in
the
non-perturbative treatment to the
$\partial_{\tau^{11}}+\partial_{\tau^{22}}$.
The  ``back-reaction'' term  can be fixed since it has to
restore the modular properties of the ${\bar \tau}^ij$ derivatives.

The physical interpretation of the summation over the ``magnetic"
charges reproduces the Euclidean space-time instanton corrections
to the couplings.

The determination of the non-perturbative effective couplings
constants (the gravitational one included) defines without any
ambiguity the non-perturbative prepotential of  the  $N=2$ effective
theory. Therefore, the knowledge of ${\cal L}^{\cal D}_{l}$
determines   at the non-perturbative level
the $N=2$ low-energy  effective  supergravity,  which includes terms
up to two derivatives.

\section{Outlook}

We have demonstrated the existence of partial spontaneous
supersymmetry
breaking in string theory, and gave several concrete examples in both
 the heterotic and type II theories.
We have studied the issue of restoration of supersymmetry, at the
classical and perturbative level.
We have further analysed the consequences of heterotic--type II
duality
valid for the $N=2$ models we presented. We have pointed out that in
the
dual theories the $N=4\to N=2$ supersymmetries may look  explicitly
broken in their perturbation theory.
This was also corroborated by our conjecture on the full
non-perturbative structure of their effective theories.
In some cases we can predict some novel non-perturbative
(non-geometric) transitions
between vacua of the type II string with (2,0) and (1,1) space-time
supersymmetry.

An analysis of the perturbative BPS states of strings,
with supersymmetry spontaneously broken $N=4\to N=2$, and the
underlying
duality structure permit us to conjecture the full non-perturbative
form of the effective field theory.
This conjecture needs to be elaborated and tested in the context of
explicit models. This will be the subject of future analysis.
\vskip .2cm
\centerline{\bf Acknowledgements}
Part of this  work was done in collaboration with E. Kiritsis.
We acknowledge useful discussions with B. Pioline and E. Verlinde.
This work is supported in part by the TMR contract
ERB--4061--PL--95--0789
in which the author is associated to ENS.

\end{document}